\documentclass[12pt,a4paper]{elsarticle}
\usepackage{ifthen}
\usepackage{bm}
\usepackage{amssymb}
\usepackage{amsfonts}
\usepackage{upgreek}
\usepackage{xspace}
\usepackage{amsmath}

\usepackage{lineno}  

\usepackage{graphicx}
\usepackage{subfigure}

\newboolean{pdflatex}
\setboolean{pdflatex}{true} 

\newboolean{articletitles}
\setboolean{articletitles}{true}
\newboolean{uprightparticles}
\setboolean{uprightparticles}{false} 

\usepackage{hyperref}
\usepackage[all]{hypcap}

\def\lhcb {LHCb\xspace}                
\def\PJ      {\ensuremath{J}\xspace}                 
\def\Pc      {\ensuremath{c}\xspace}                                  
\def\Pmu         {\ensuremath{\mu}\xspace}                 
\def\Ppi         {\ensuremath{\pi}\xspace}                 
\def\Pchi        {\ensuremath{\chi}\xspace}                 
\def\Ppsi        {\ensuremath{\psi}\xspace}                 
\def\mup        {\ensuremath{\Pmu^+}\xspace}
\def\mun        {\ensuremath{\Pmu^-}\xspace} 
\def\cquark    {\ensuremath{\Pc}\xspace}
\def\cquarkbar {\ensuremath{\overline \cquark}\xspace}
\def\ccbar     {\ensuremath{\cquark\cquarkbar}\xspace}
\def\pion  {\ensuremath{\Ppi}\xspace}
\def\piz   {\ensuremath{\pion^0}\xspace}
\def\pim   {\ensuremath{\pion^-}\xspace}
\def\jpsi     {\ensuremath{{\PJ\mskip -3mu/\mskip -2mu\Ppsi\mskip 2mu}}\xspace}
\def\psitwos  {\ensuremath{\Ppsi{(2S)}}\xspace}
\def\chiczero {\ensuremath{\Pchi_{\cquark 0}}\xspace}
\def\chicone  {\ensuremath{\Pchi_{\cquark 1}}\xspace}
\def\chictwo  {\ensuremath{\Pchi_{\cquark 2}}\xspace}
\def\chic  {\ensuremath{\Pchi_{c}}\xspace}
\def\BF         {{\ensuremath{\cal B}\xspace}}
\def\BR         {\BF}
\newcommand{\tev}{\ensuremath{\mathrm{\,Te\kern -0.1em V}}\xspace}
\newcommand{\gevc}{\ensuremath{{\mathrm{\,Ge\kern -0.1em V\!/}c}}\xspace}
\newcommand{\mevc}{\ensuremath{{\mathrm{\,Me\kern -0.1em V\!/}c}}\xspace}
\newcommand{\gevcc}{\ensuremath{{\mathrm{\,Ge\kern -0.1em V\!/}c^2}}\xspace}
\newcommand{\mevcc}{\ensuremath{{\mathrm{\,Me\kern -0.1em V\!/}c^2}}\xspace}
\def\invpb {\ensuremath{\mbox{\,pb}^{-1}}\xspace}
\def\ps   {\ensuremath{{\rm \,ps}}\xspace}
\def\mrad{\ensuremath{\rm \,mrad}\xspace}
\def\evtgen     {\mbox{\textsc{EvtGen}}\xspace}
\def\pythia     {\mbox{\textsc{Pythia}}\xspace}
\def\geant      {\mbox{\textsc{Geant4}}\xspace}

\def\photos     {\mbox{\textsc{Photos}}\xspace}
\def\chigen     {\mbox{\textsc{ChiGen}}\xspace}
\newcommand{\ie}{\mbox{\itshape i.e.}}

\newcommand{\Figure}{Figure}
\newcommand{\aFigure}{Fig.}

\newcommand{\aTable}{Table}

\newcommand{\aSection}{Sect.}

\newcommand{\aEquation}{Eq.}

\newcommand{\aReference}{Ref.}

\newcommand{\MeVc}{\ensuremath{\mathrm{MeV}/c}}
\newcommand{\MeVcc}{\ensuremath{\mathrm{MeV}/c^{2}}}
\newcommand{\GeV}{\ensuremath{\mathrm{GeV}}}
\newcommand{\GeVc}{\ensuremath{\mathrm{GeV}/c}}
\newcommand{\GeVcc}{\ensuremath{\mathrm{GeV}/c^{2}}}
\newcommand{\TeV}{\ensuremath{\mathrm{TeV}}}

\newcommand{\pp}{\ensuremath{pp}}
\newcommand{\ppbar}{\ensuremath{p\bar{p}}}
\newcommand{\pA}{\ensuremath{p\mathrm{A}}}
\newcommand{\Jpsi}{\jpsi}
\newcommand{\Chic}{\Pchi\xspace} 

\newcommand{\PsiTwoS}{\psitwos}
\newcommand{\ChicJ}{\ensuremath{\chi_{cJ}}}
\newcommand{\ChicJOneP}{\ensuremath{\ChicJ(1P)}}

\newcommand{\ChicZero}{\ensuremath{\chi_{c0}}}

\newcommand{\ChicOne}{\ensuremath{\chi_{c1}}}

\newcommand{\ChicTwo}{\ensuremath{\chi_{c2}}}

\newcommand{\bhadron}{\ensuremath{b}-hadron}

\newcommand{\myPV}{\ensuremath{\mathrm{PV}}}
\newcommand{\pT}{\ensuremath{p_{\mathrm{T}}}}
\newcommand{\pTJpsi}{\ensuremath{\pT^{\Jpsi}}}

\newcommand{\pTGamma}{\ensuremath{\pT^{\gamma}}}
\newcommand{\pGamma}{\ensuremath{p^{\gamma}}}
\newcommand{\CLgamma}{\ensuremath{\mathrm{CL}_{\gamma}}}

\newcommand{\RapidityJpsi}{\ensuremath{y^{\Jpsi}}}

\newcommand{\tz}{\ensuremath{t_{z}}}
\newcommand{\SqrtS}{\ensuremath{\sqrt{s}\myop{=}\myvalue{7}{\TeV}}}

\newcommand{\MuMu}{\ensuremath{\mup\,\mun}}
\newcommand{\JpsiToMuMu}{\ensuremath{\Jpsi\,\rightarrow\,\mup\,\mun}}

\newcommand{\ChicToJpsiGamma}{\ensuremath{\Chic\,\rightarrow\,\Jpsi\,\gamma}}
\newcommand{\ChicZeroToJpsiGamma}{\ensuremath{\ChicZero\,\rightarrow\,\Jpsi\,\gamma}}
\newcommand{\ChicOneToJpsiGamma}{\ensuremath{\ChicOne\,\rightarrow\,\Jpsi\,\gamma}}
\newcommand{\ChicTwoToJpsiGamma}{\ensuremath{\ChicTwo\,\rightarrow\,\Jpsi\,\gamma}}

\newcommand{\rChicTwoToChicOne}{\ensuremath{\frac{\sigma(\ChicTwo)}{\sigma(\ChicOne)}}}

\newcommand{\srChicTwoToChicOne}{\ensuremath{\sigma(\ChicTwo)\,/\,\sigma(\ChicOne)}}

\newcommand{\DeltaM}{\ensuremath{\Delta M}}

\newcommand{\pTJpsiRange}{\ensuremath{2\myop{<}\pTJpsi\myop{<}\myvalue{15}{\GeVc}}}
\newcommand{\pTJpsiRangeIntSample}{\ensuremath{3\myop{<}\pTJpsi\myop{<}\myvalue{15}{\GeVc}}}

\newcommand{\yRange}{\ensuremath{2.0\myop{<}\RapidityJpsi\myop{<}4.5}}

\newcommand{\FitFunc}{\ensuremath{\mathcal{F}}}
\newcommand{\FitFuncSig}{\ensuremath{\mathcal{F}_{\mathrm{sig}}^{J}}}
\newcommand{\FitFuncBkg}{\ensuremath{\mathcal{F}_{\mathrm{bgd}}}}

\providecommand{\SigmaRes}[1]{\ensuremath{\sigma_{\mathrm{res}}^{#1}}}
\providecommand{\SigmaResRatio}[2]{\ensuremath{\SigmaRes{#1}\myop{/}\SigmaRes{#2}}}

\newcommand{\LHCb}{LHCb}

\newcommand{\myapprox}[1]{\ensuremath{\sim{#1}}}
\newcommand{\mypma}[1]{\ensuremath{\myopone{\pm}{#1}}}
\newcommand{\myapmb}[2]{\ensuremath{{#1}\myop{\pm}{#2}}}

\newcommand{\myop}[1]{\ensuremath{\,{#1}\,}}
\newcommand{\myopone}[1]{\ensuremath{{#1}\,}}
\newcommand{\myrange}[2]{\ensuremath{{#1}\myop{-}{#2}}}
\newcommand{\myTabRange}[2]{\ensuremath{{#1}\!-\!{#2}}}
\newcommand{\myvalue}[2]{\mbox{\ensuremath{{#1}\:{#2}}}} 

\renewcommand{\MeVc}{\mevc}
\renewcommand{\MeVcc}{\mevcc}
\renewcommand{\GeVc}{\gevc}
\renewcommand{\GeVcc}{\gevcc}
\renewcommand{\TeV}{\tev}

\renewcommand{\Jpsi}{\jpsi}
\renewcommand{\Chic}{\chic}
\renewcommand{\ChicZero}{\chiczero}
\renewcommand{\ChicOne}{\chicone}
\renewcommand{\ChicTwo}{\chictwo}
\renewcommand{\LHCb}{\lhcb}

\usepackage{multirow}
\usepackage{mciteplus} 

\begin{document}


\begin{titlepage}
\pagenumbering{roman}

\vspace*{-1.5cm}
\centerline{\large EUROPEAN ORGANIZATION FOR NUCLEAR RESEARCH (CERN)}
\vspace*{1.5cm}
\hspace*{-0.5cm}
\begin{tabular*}{\linewidth}{lc@{\extracolsep{\fill}}r}
\ifthenelse{\boolean{pdflatex}}
{\vspace*{-2.7cm}\mbox{\!\!\!\includegraphics[width=.14\textwidth]{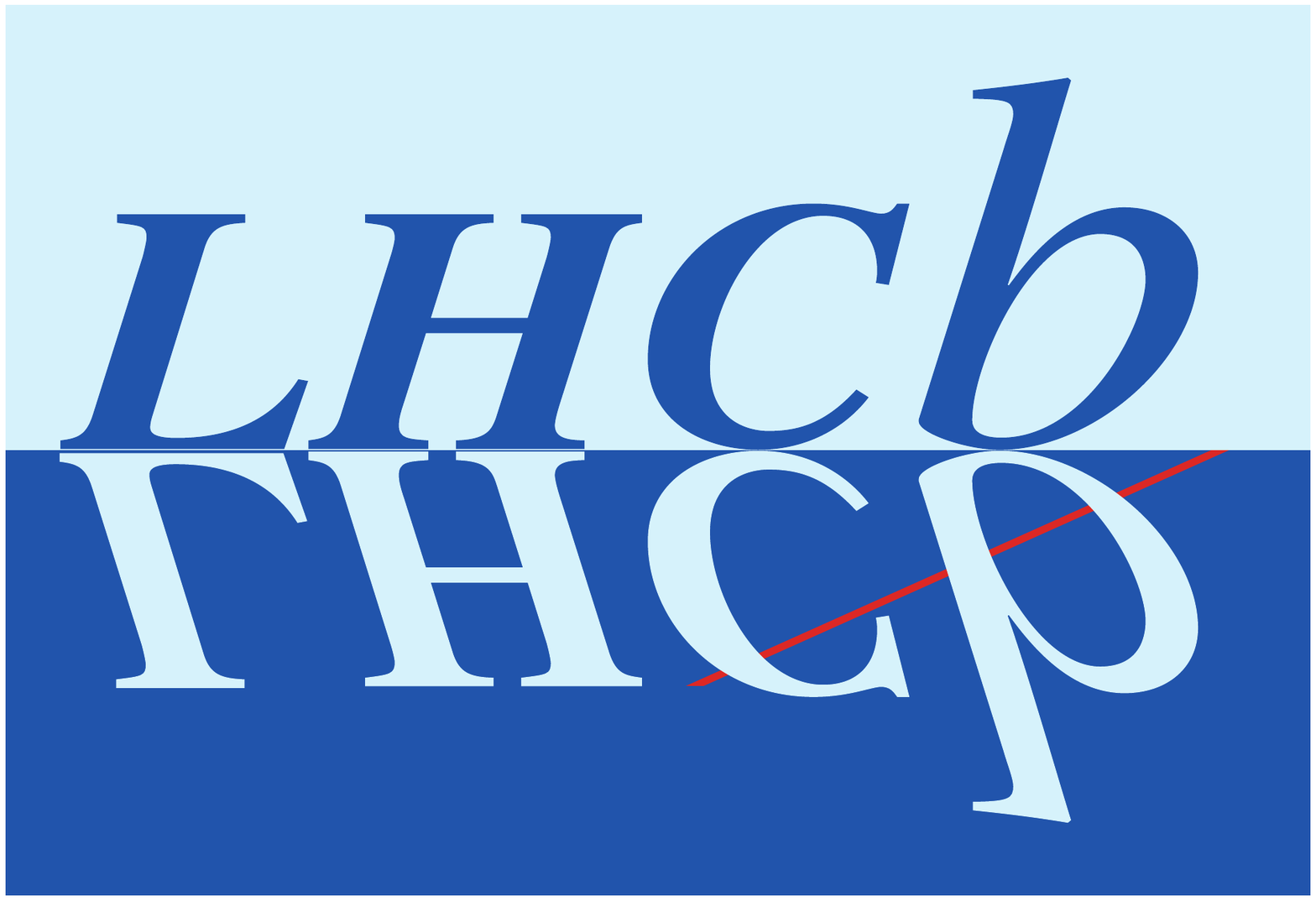}} & &}%
{\vspace*{-1.2cm}\mbox{\!\!\!\includegraphics[width=.12\textwidth]{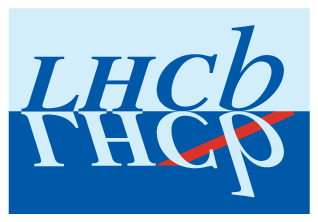}} & &}%
\\
 & & CERN-PH-EP-2011-227 \\  
 & & LHCb-PAPER-2011-019 \\  
 & & \today \\ 
 & & \\
\end{tabular*}

\vspace*{4.0cm}

{\bf\boldmath\huge
\begin{center}
  Measurement of the cross-section ratio ${\srChicTwoToChicOne}$
for prompt ${\Chic}$ production at ${\SqrtS}$
\end{center}
}

\vspace*{2.0cm}

\begin{center}
The LHCb Collaboration
\footnote{Authors are listed on the following pages.}
\end{center}

\vspace{\fill}

\begin{abstract}
  \noindent 
The prompt production of the charmonium $\chi_{c1}$ and 
$\chi_{c2}$ mesons has been
studied in proton-proton collisions at the Large Hadron Collider at a
centre-of-mass energy of 
$\sqrt{s}=7$~TeV. The $\chi_c$ mesons are identified through
their decays 
$\chi_c\,\rightarrow\,J/\psi\,\gamma$
with 
$J/\psi\,\rightarrow\,\mu^+\,\mu^-$
using 36~$\mathrm{pb^{-1}}$
of data collected by the LHCb detector in 2010.  The ratio of the prompt
production cross-sections for the two $\chi_c$ spin states, 
$\sigma(\chi_{c2})\,/\,\sigma(\chi_{c1})$,
has been determined as a function of the $J/\psi$ transverse momentum, 
$p_{\mathrm{T}}^{J/\psi}$,
in the range from 2 to 15~GeV/$c$. The results are in agreement with
the next-to-leading order non-relativistic QCD model at high $p_{\mathrm{T}}^{J/\psi}$ 
and lie consistently above the pure leading-order colour singlet prediction.
\end{abstract}

\vspace*{1.0cm}

\begin{center}
\it{Submitted to Phys. Lett. B}
\end{center}

\vspace{\fill}

\end{titlepage}

\maketitle

\newpage
\setcounter{page}{2}
\mbox{~}
\newpage

\bigskip
\centerline{\large\bf{LHCb Collaboration}}
\begin{flushleft}
\small
R.~Aaij$^{23}$, 
C.~Abellan~Beteta$^{35,n}$, 
B.~Adeva$^{36}$, 
M.~Adinolfi$^{42}$, 
C.~Adrover$^{6}$, 
A.~Affolder$^{48}$, 
Z.~Ajaltouni$^{5}$, 
J.~Albrecht$^{37}$, 
F.~Alessio$^{37}$, 
M.~Alexander$^{47}$
% ...

{\footnotesize \it
% ...
$ ^{5}$Clermont Universit\'{e}, Universit\'{e} Blaise Pascal, CNRS/IN2P3, LPC, Clermont-Ferrand, France\\
$ ^{6}$CPPM, Aix-Marseille Universit\'{e}, CNRS/IN2P3, Marseille, France\\
% ...
$ ^{23}$Nikhef National Institute for Subatomic Physics, Amsterdam, The Netherlands\\
% ...
$ ^{35}$Universitat de Barcelona, Barcelona, Spain\\
$ ^{36}$Universidad de Santiago de Compostela, Santiago de Compostela, Spain\\
$ ^{37}$European Organization for Nuclear Research (CERN), Geneva, Switzerland\\
% ...
$ ^{42}$H.H. Wills Physics Laboratory, University of Bristol, Bristol, United Kingdom\\
% ...
$ ^{47}$School of Physics and Astronomy, University of Glasgow, Glasgow, United Kingdom\\
$ ^{48}$Oliver Lodge Laboratory, University of Liverpool, Liverpool, United Kingdom\\
% ...
$ ^{n}$LIFAELS, La Salle, Universitat Ramon Llull, Barcelona, Spain\\
% ...
}
\clearpage
\end{flushleft}

\cleardoublepage

\pagestyle{plain} 
\setcounter{page}{1}
\pagenumbering{arabic}




\section{Introduction}
\label{sec:Introduction}

Explaining heavy quarkonium production remains a challenging problem for Quantum
Chromodynamics (QCD). At the energies of the proton-proton (\pp) collisions at
the Large Hadron Collider, \ccbar pairs are expected to be produced
predominantly via Leading Order (LO) gluon-gluon interactions, followed by the
formation of the bound charmonium states. While the former can be calculated
using perturbative QCD, the latter is described by non-perturbative
models. Other, more recent, approaches make use of non-relativistic QCD
factorization (NRQCD) which assumes a combination of the colour-singlet (CS) and
colour-octet (CO) \ccbar and soft gluon exchange for the production of the final
bound state~\cite{Bodwin:1994jh}. 
To describe previous experimental data, it was
found to be necessary 
to include Next-to-Leading Order (NLO) QCD corrections for the
description of charmonium production~\cite{Campbell:2007ws,Ma:2010vd}.

The study of the production of $P$-wave charmonia \ChicJOneP, with
\ensuremath{J\myop{=}0,1,2}, is important, since these resonances give
substantial feed-down contributions to the prompt \Jpsi\ production through
their radiative decays \ChicToJpsiGamma\ and can have significant impact on the
measurement of the \Jpsi\ polarisation. Furthermore, the ratio of the production
rate of \ChicTwo\ to that of \ChicOne\ is interesting because it is sensitive to
the CS and CO production mechanisms.

Measurements of \Chic\ production and the relative amounts of the \ChicOne\ and
\ChicTwo\ spin states, have previously been made using different particle beams
and energies~\cite{Lemoigne:1982jc,Abt:2008ed,Abulencia:2007bra}. In this
Letter, we report a measurement from the \LHCb\ experiment of the ratio of the
prompt cross-sections for the two \Chic\ spin states, \srChicTwoToChicOne, as a
function of the \Jpsi\ transverse momentum in the range \pTJpsiRange\ and in the
rapidity range \yRange. The \Chic\ candidates are reconstructed through their
radiative decay \ChicToJpsiGamma, with \JpsiToMuMu, using a data sample with an
integrated luminosity of \myvalue{36}{\invpb} collected during 2010. In this
Letter, prompt production of \Chic\ refers to \Chic\ mesons that are produced at
the interaction point and do not arise from the decay of a \bhadron. The sample
therefore includes \Chic\ from the decay of short-lived resonances, such as
\PsiTwoS, which are also produced at the interaction point. All three \ChicJ\
states are considered in the analysis. Since the \ChicZeroToJpsiGamma\ branching
fraction is \myapprox{30} (17) times smaller than that of the 
\ChicOne\ (\ChicTwo), 
the yield of \ChicZero\ is not significant. The measurements extend
the \pTJpsi\ coverage with respect to previous experiments.


\section{\LHCb\ detector and selection requirements}
\label{sec:Detector}

The \LHCb\ detector~\cite{Alves:2008zz} is a single-arm forward spectrometer
with an angular coverage from approximately \myvalue{10}{\mrad} to
\myvalue{300}{\mrad} (\myvalue{250}{\mrad}) in the bending (non-bending)
plane. The detector consists of a vertex detector (VELO), a dipole magnet, a
tracking system, two ring-imaging Cherenkov (RICH) detectors, a calorimeter
system and a muon system.

Of particular importance in this measurement are the calorimeter and muon
systems. The calorimeter consists of a scintillating pad detector (SPD) and a
pre-shower, followed by electromagnetic (ECAL) and hadronic calorimeters. The
SPD and pre-shower are designed to distinguish between signals from photons and
electrons. The ECAL is constructed from scintillating tiles interleaved with
lead tiles. Muons are identified using hits in detectors interleaved with iron
filters.

The signal simulation sample used for this analysis was generated using the
\pythia~\ensuremath{6.4} generator~\cite{Sjostrand:2006za} 
configured with the parameters detailed 
in \aReference~\cite{LHCb-PROC-2010-056}.
The \evtgen~\cite{Lange:2001uf}, \photos~\cite{Barberio:1993qi} 
and \geant~\cite{Agostinelli:2002hh} 
packages were used to decay unstable particles, 
generate QED radiative corrections and 
simulate interactions in the detector, respectively.
The sample consists of events in which at least one \JpsiToMuMu\ decay
takes place with no constraint on the production mechanism.

The trigger consists of a hardware stage followed by a software stage which
applies a full event reconstruction. For this analysis the trigger selects a
pair of oppositely charged muon candidates, where either one of the muons has a
transverse momentum \ensuremath{\pT\myop{>}\myvalue{1.8}{\GeVc}} or one of the
pair has \ensuremath{\pT\myop{>}\myvalue{0.56}{\GeVc}} and the other has
\ensuremath{\pT\myop{>}\myvalue{0.48}{\GeVc}}. The invariant mass of the
candidates is required to be greater than \myvalue{2.9}{\GeVcc}. The photons are
not involved in the trigger decision for this analysis.

Photons are identified and reconstructed using the calorimeter and tracking
systems. The identification algorithm provides an estimator for the hypothesis
that a calorimeter cluster originates from a photon. This is a likelihood-based
estimator constructed from variables that rely on calorimeter and tracking
information. For example, in order to reduce the electron background, candidate
photon clusters are required not to be matched to a track extrapolated into the
calorimeter. For each photon candidate a likelihood (\CLgamma) is calculated
based on simulated signal and background samples. The photons identified by the
calorimeter and used in this analysis can be classified as two types: those that
have converted in the material after the dipole magnet and those that have
not. Converted photons are identified as clusters in the ECAL with correlated
activity in the SPD. In order to account for the different energy resolutions of
the two types of photons, the analysis is performed separately for converted and
non-converted photons and the results combined as described in
\aSection~\ref{sec:ExpMethod}. Photons that convert before the magnet require a
different analysis strategy and are not considered here. The photons used to
reconstruct the \Chic\ candidates are required to have a transverse momentum
\ensuremath{\pTGamma\myop{>}\myvalue{650}{\MeVc}}, a momentum
\ensuremath{\pGamma\myop{>}\myvalue{5}{\GeVc}} and a likelihood
\ensuremath{\CLgamma\myop{>}0.5}.

The muon and \Jpsi\ identification criteria are identical to those used in
\aReference~\cite{Aaij:2011jh}: each track must be identified as a muon with
\ensuremath{\pT\myop{>}700\MeVc} and a quality of the track fit
\ensuremath{\chi^{2}/\mathrm{ndf}\myop{<}4}, where \ensuremath{\mathrm{ndf}} is
the number of degrees of freedom. The two muons must originate from a common
vertex with a probability of the vertex fit \ensuremath{\myopone{>}0.5\%}. In
addition, in this analysis the \MuMu\ invariant mass is required to be in the
range \myrange{3062}{\myvalue{3120}{\MeVcc}}. The \Jpsi\ pseudo-decay time, \tz,
is used to reduce the contribution from non-prompt decays, by requiring
\ensuremath{\tz\myop{=}(z_{\Jpsi}\myop{-}z_{\myPV})M_{\Jpsi}\myop{/}p_{z}\myop{<}\myvalue{0.1}{\ps}},
where \ensuremath{M_{\Jpsi}} is the reconstructed dimuon invariant mass,
\ensuremath{z_{\Jpsi}\myop{-}z_{\myPV}} is the \ensuremath{z} separation of the
reconstructed production (primary) and decay vertices of the dimuon, 
and \ensuremath{p_{z}} is
the \ensuremath{z}-component of the dimuon momentum with the \ensuremath{z}-axis
parallel to the beam line. Simulation studies show that, with this requirement
applied, the remaining fraction of \Chic\ from \bhadron\ decays is about
\ensuremath{0.1\%}. This introduces an uncertainty much smaller than any of the
other systematic or statistical uncertainties evaluated in this analysis and is
not considered further.

\begin{figure}
  \begin{center}
    \subfigure{
      \ifthenelse{\boolean{pdflatex}}{
	\includegraphics*[width=0.45\textwidth]{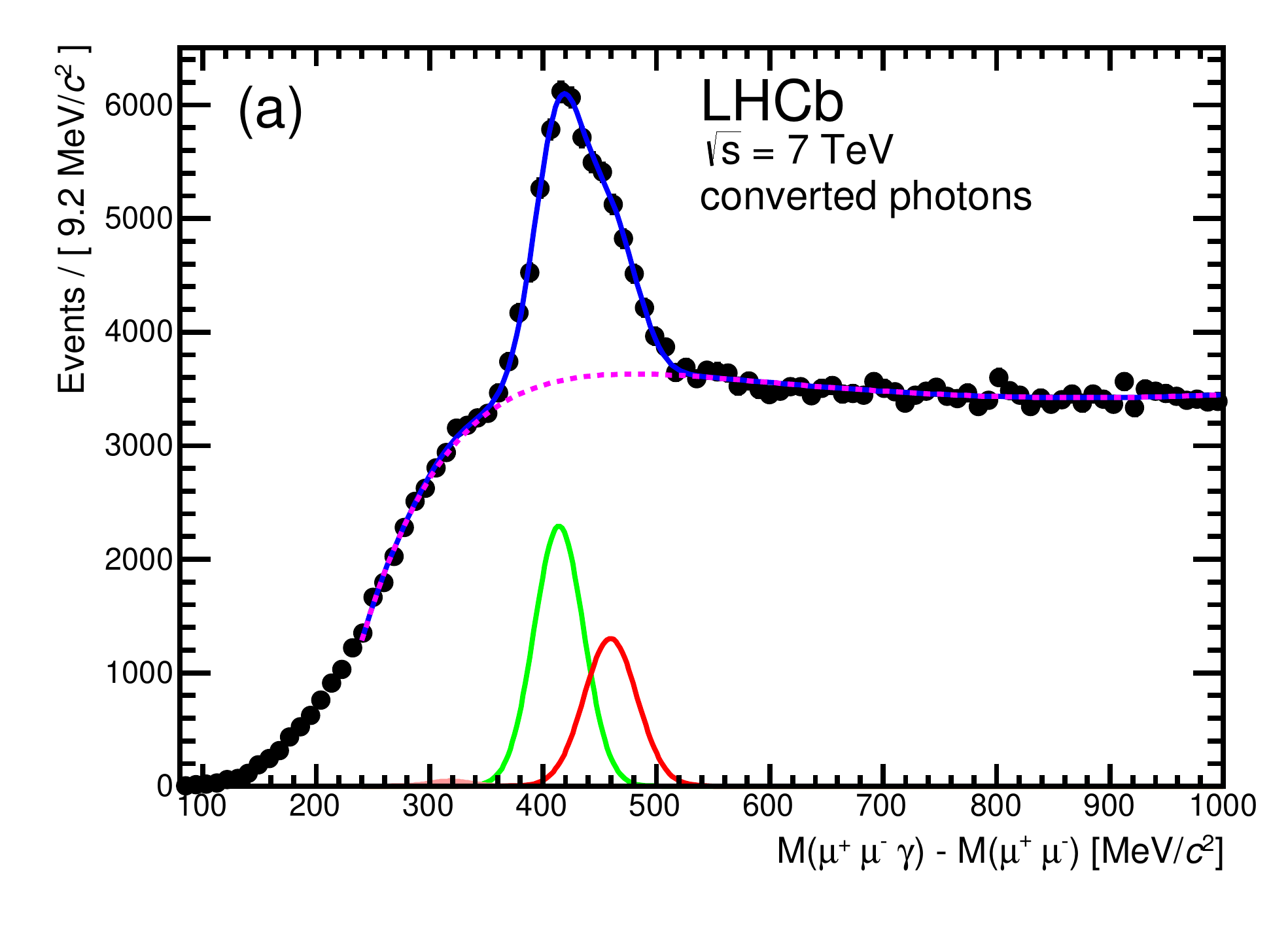}
      }{
	\includegraphics*[width=0.45\textwidth]{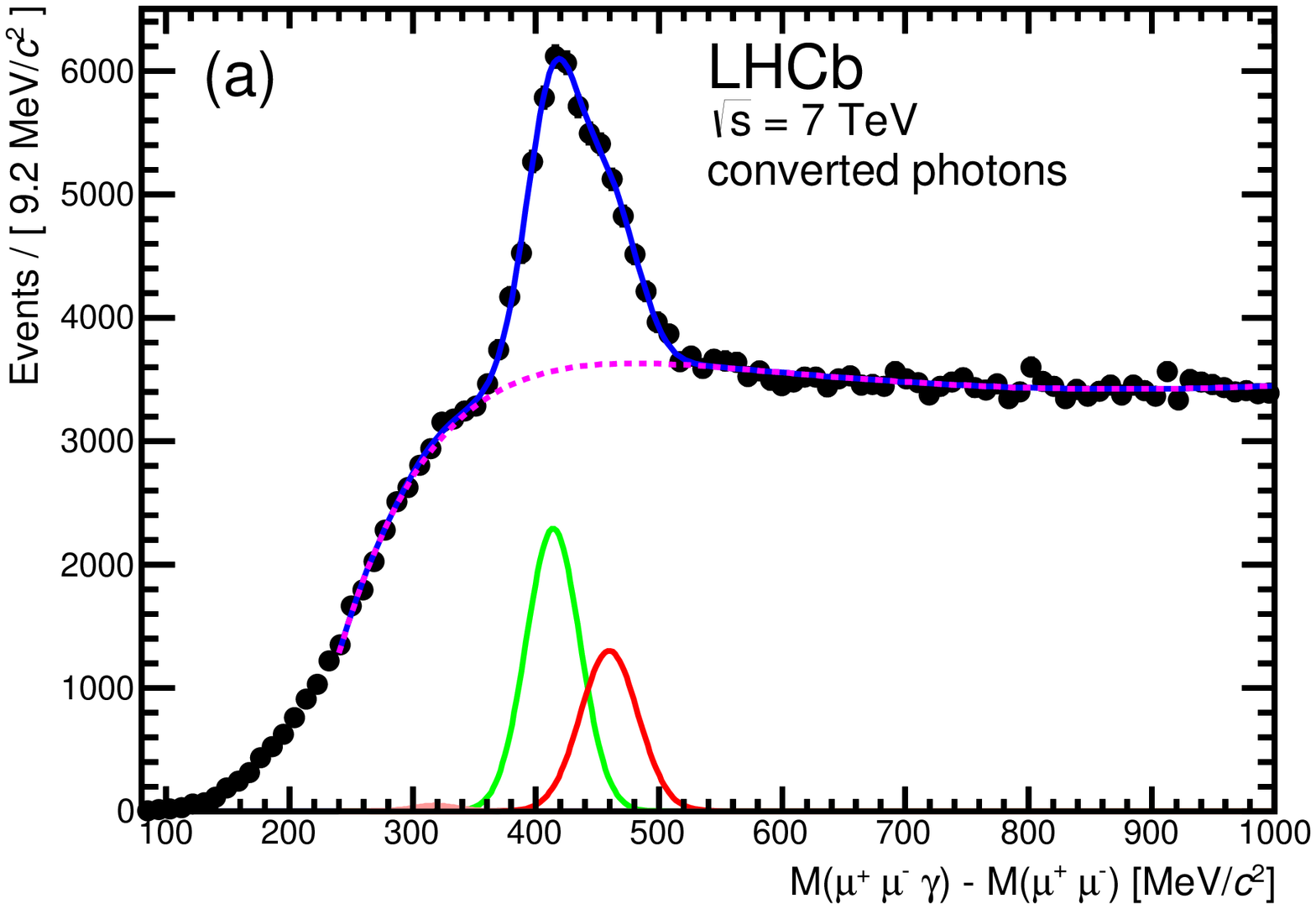}}}
    \subfigure{
      \ifthenelse{\boolean{pdflatex}}{
	\includegraphics*[width=0.45\textwidth]{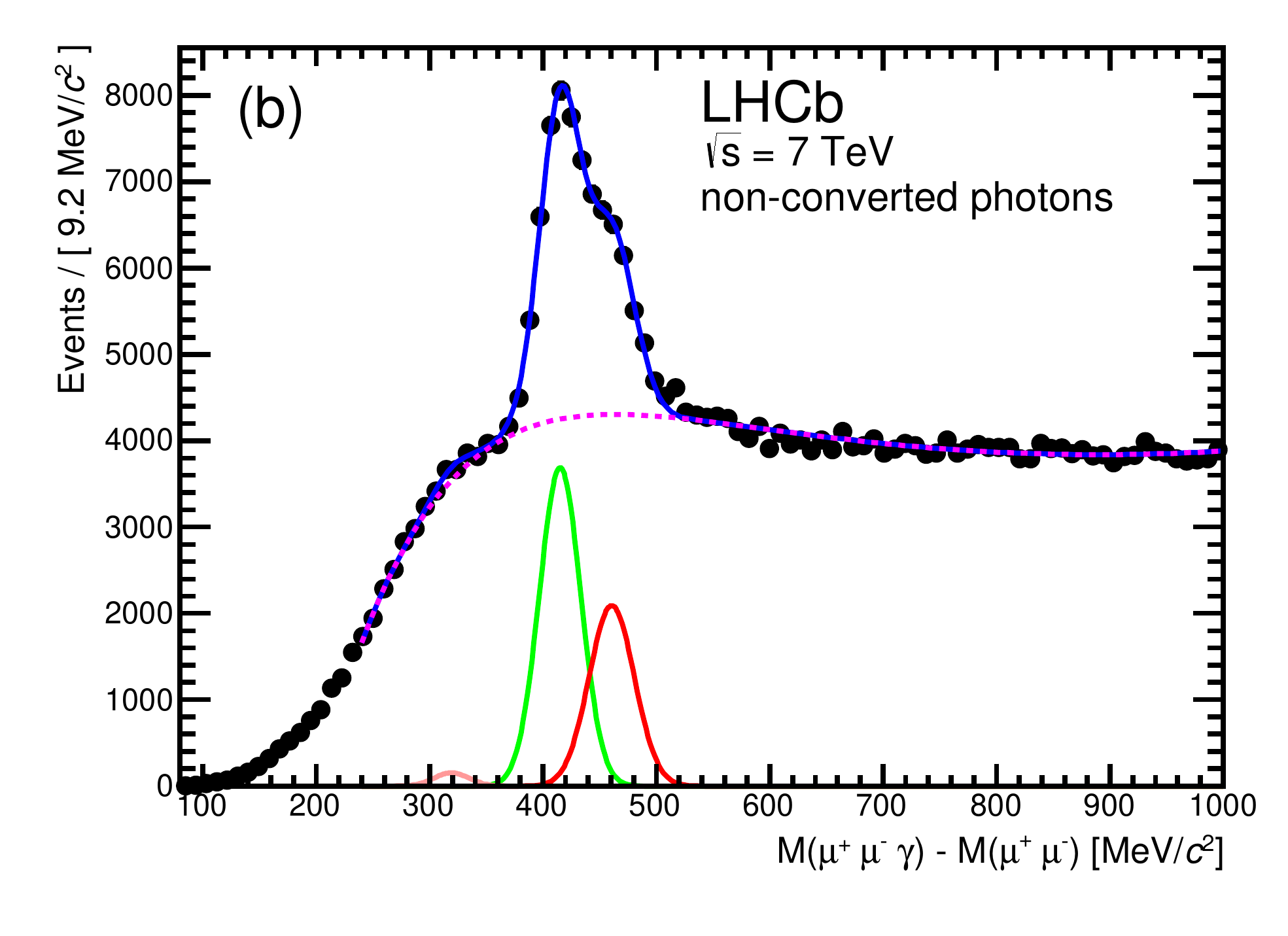}
      }{
	\includegraphics*[width=0.45\textwidth]{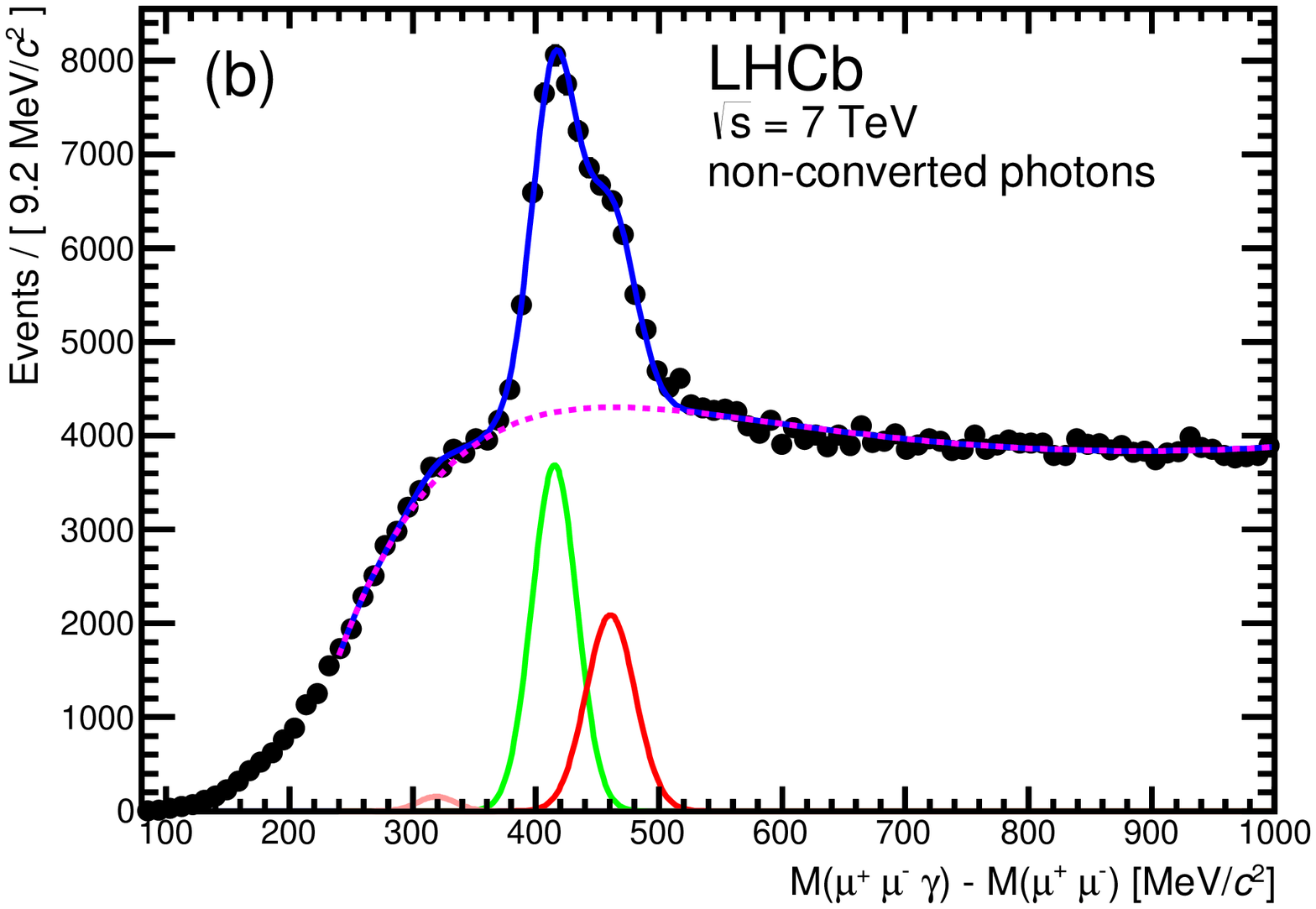}}}
    \caption{
      \label{fig:MassPlots}
      \small{Distribution of
	\ensuremath{\DeltaM\myop{=}M\left(\mup\,\mun\,\gamma\right)\myop{-}M\left(\mup\,\mun\right)}
	for selected candidates with \pTJpsiRangeIntSample\ for (a) converted
	and (b) non-converted photons. The lower solid curves correspond to the
	\ChicZero, \ChicOne\ and \ChicTwo\ peaks from left to right,
	respectively (the \ChicZero\ peak is barely visible). The background
	distribution is shown as a dashed curve. The upper solid curve corresponds to the
	overall fit function.}}
  \end{center}
\end{figure}

In the data, the average \Chic\ candidate multiplicity per selected event is
$1.3$ and the percentage of events with more than one genuine \Chic\ candidate
(composed of a unique \Jpsi\ and photon) is estimated to be $0.23\%$ from the
simulation. All \Chic\ candidates are considered for further analysis. The mass
difference,
\ensuremath{\DeltaM\myop{=}M\left(\mup\,\mun\,\gamma\right)\myop{-}M\left(\mup\,\mun\right)},
of the selected candidates is shown in \aFigure~\ref{fig:MassPlots} for the
converted and non-converted samples; the overlaid fits are described in
\aSection~\ref{sec:ExpMethod}.  


\section{Experimental method}
\label{sec:ExpMethod}

The production cross-section ratio of the \ChicTwo\ and \ChicOne\ states is
measured as
\begin{gather}
    \label{e:ChiC1_ChiC2_ratio}
    \rChicTwoToChicOne\myop{=}\frac{N_{\ChicTwo}}{N_{\ChicOne}}\myop{\cdot}\frac{\epsilon^{\ChicOne}}{\epsilon^{\ChicTwo}}\myop{\cdot}\frac{\BR(\ChicOneToJpsiGamma)}{\BR(\ChicTwoToJpsiGamma)},
\end{gather}
where \ensuremath{\BR(\ChicOneToJpsiGamma)} and
\ensuremath{\BR(\ChicTwoToJpsiGamma)} are the \ChicOne\ and \ChicTwo\ branching
fractions to the final state \ensuremath{\Jpsi\,\gamma}, and
\begin{gather}
    \label{e:ChiC1_ChiC2_EffRatio}
    \frac{\epsilon^{\ChicOne}}{\epsilon^{\ChicTwo}}\myop{=}\frac{\epsilon^{\ChicOne}_{\Jpsi}\,\epsilon^{\ChicOne}_{\gamma}\,\epsilon^{\ChicOne}_{\mathrm{sel}}}{\epsilon^{\ChicTwo}_{\Jpsi}\,\epsilon^{\ChicTwo}_{\gamma}\,\epsilon^{\ChicTwo}_{\mathrm{sel}}},
\end{gather}
where \ensuremath{\epsilon^{\ChicJ}_{\Jpsi}} is the efficiency to trigger,
reconstruct and select a \Jpsi\ from a \ChicJ\ decay,
\ensuremath{\epsilon^{\ChicJ}_{\gamma}} is the efficiency to reconstruct and
select a photon from a \ChicJ\ decay and
\ensuremath{\epsilon^{\ChicJ}_{\mathrm{sel}}} is the efficiency to subsequently
select the \ChicJ\ candidate.

Since the mass difference between the \ChicOne\ and \ChicTwo\ states is
\myvalue{\myapmb{45.54}{0.11}}{\MeVcc}, the signal peaks cannot be separately
isolated using the calorimeter information. An unbinned maximum likelihood fit
to the \DeltaM\ mass difference distribution is performed to obtain the three
\ensuremath{N_{\ChicJ}} yields simultaneously. The determination of the efficiency terms in
\aEquation~\ref{e:ChiC1_ChiC2_EffRatio} is described in
\aSection~\ref{sec:Efficiencies}.

The signal mass distribution is parametrised using three Gaussian functions
(\FitFuncSig\ for \ensuremath{J\myop{=}0,1,2}). The combinatorial background is
described by
\begin{align}
  \FitFuncBkg\myop{=}x^{a}\!\left(1-e^{\frac{m_{0}}{c}\left(1-x\right)}\right)+b\left(x-1\right),
\end{align}
where \ensuremath{x\myop{=}\DeltaM\myop{/}m_{0}} and $m_0$, $a$, $b$ and $c$ are
free parameters.

A possible source of background from partially reconstructed decays is due to
\ensuremath{\PsiTwoS\myop{\rightarrow}\Jpsi\,\piz\,\piz} decays where the \Jpsi\
and a photon from one of the neutral pions are reconstructed and selected as a
\Chic\ candidate. Simulation studies show that the expected yield is
\myapprox{0.1\%} of the signal yield and this background is therefore neglected
for this analysis.

The overall fit function is 
\begin{gather}
  \label{e:FitFunction}
  \FitFunc\myop{=}\sum_{J=0}^2 f_{\ChicJ}\,\FitFuncSig\myop{+}\left[1\myop{-}\sum_{J=0}^2 f_{\ChicJ}\right]\FitFuncBkg,
\end{gather}
where \ensuremath{f_{\ChicJ}} are the signal fractions. The mass differences
between the \ChicOne\ and \ChicTwo\ states and the \ChicOne\ and \ChicZero\
states are fixed to the values from \aReference~\cite{Nakamura:2010zzi}. The
mass resolutions for the \Chic\ states, \SigmaRes{\ChicJ}, are given by the
widths of the Gaussian functions for each state. The ratios of the mass
resolutions, \SigmaResRatio{\ChicTwo}{\ChicOne} and
\SigmaResRatio{\ChicZero}{\ChicOne}, are taken from simulation. The value of
\SigmaResRatio{\ChicTwo}{\ChicOne} is consistent with the value measured from
data, fitting in a reduced \DeltaM\ range and with a simplified background
parametrisation.

With the mass differences and the ratio of the mass resolutions fixed, a fit is
performed to the data in the range \pTJpsiRangeIntSample, in order to determine
the \ChicOne\ mass resolution \SigmaRes{\ChicOne}. This range is chosen because
the background has a different shape in the \pTJpsi\ bin
\myvalue{\myrange{2}{3}}{\GeVc} and is not well described by \FitFuncBkg\ when
combined with the rest of the sample. Simulation studies show that the signal
parameters for the \ChicJ\ states in the \pTJpsi\ bin
\myvalue{\myrange{2}{3}}{\GeVc} are consistent with the parameters in the rest
of the sample. 
The distributions of \DeltaM\ for the fits to the converted and
non-converted candidates are shown in \aFigure~\ref{fig:MassPlots}. The mass
resolution, \SigmaRes{\ChicOne}, is measured to be
\myvalue{\myapmb{21.8}{0.8}}{\MeVcc} and \myvalue{\myapmb{18.3}{0.4}}{\MeVcc}
for converted and non-converted candidates respectively. The corresponding
values in the simulation are \myvalue{\myapmb{19.0}{0.2}}{\MeVcc} and
\myvalue{\myapmb{17.5}{0.1}}{\MeVcc} and show a weak dependence of
\SigmaRes{\ChicOne} on \pTJpsi\ which is accounted for in the systematic
uncertainties.

In order to measure the \Chic\ yields, the fit is then performed in bins of
\pTJpsi\ in the range \pTJpsiRange. For each \pTJpsi\ bin, the mass differences,
the ratio of the mass resolutions and \SigmaRes{\ChicOne} are fixed as described
above. In total, there are eight free parameters for each fit in each bin in
\pTJpsi\ and the results are summarized in \aTable~\ref{tab:Yields};
the fit \ensuremath{\chi^2/\mathrm{ndf}} for the converted and 
non-converted samples is good in all bins. 
The total
observed yields of \ChicZero, \ChicOne\ and \ChicTwo\ are
\ensuremath{\myapmb{820}{650}}, \ensuremath{\myapmb{38\,630}{550}} and
\ensuremath{\myapmb{26\,110}{620}}, respectively, calculated from the signal
fractions \ensuremath{f_{\ChicJ}} and the number of candidates in the
sample. The raw \Chic\ yields for converted and non-converted candidates are
combined, corrected for efficiency (as described in
\aSection~\ref{sec:Efficiencies}) and the cross-section ratio is determined using
\aEquation~\ref{e:ChiC1_ChiC2_ratio}.
\begin{table*}[t] \small
  \renewcommand{\arraystretch}{1.1}
  \caption{
    \label{tab:Yields}
    \small{Signal \Chic\ yields and fit quality from the fit to the converted and non-converted candidates in each \pTJpsi\ bin.}}
  \begin{center}
    \begin{tabular}{|c|c|c|c|c|c|c|} \hline
      \multirow{3}{*}{\pTJpsi\ (\GeVc)} & \multicolumn{3}{|c|}{Converted photons} & \multicolumn{3}{|c|}{Non-converted photons}\\ \cline{2-7}
      & \ChicOne\ yield & \ChicTwo\ yield & \ensuremath{\chi^{2}\myop{/}\mathrm{ndf}} 
      & \ChicOne\ yield & \ChicTwo\ yield & \ensuremath{\chi^{2}\myop{/}\mathrm{ndf}}\\ \hline

      \myrange{2}{3} & \myapmb{3120}{248} & \myapmb{2482}{301} & 0.91 
                     & \myapmb{4080}{246} & \myapmb{3927}{280} & 1.02 \\
      \myrange{3}{4} & \myapmb{3462}{224} & \myapmb{3082}{249} & 0.81 
                     & \myapmb{4919}{183} & \myapmb{3443}{207} & 1.02 \\
      \myrange{4}{5} & \myapmb{3235}{146} & \myapmb{1769}{174} & 1.03 
                     & \myapmb{4497}{134} & \myapmb{2718}{143} & 1.08 \\
      \myrange{5}{6} & \myapmb{2476}{110} & \myapmb{1443}{121} & 0.84 
                     & \myapmb{3203}{105} & \myapmb{1999}{107} & 1.45 \\
      \myrange{6}{7} & \myapmb{1497}{80}  & \myapmb{736}{89}   & 1.05 
                     & \myapmb{1946}{78}  & \myapmb{1338}{83}  & 0.78 \\
      \myrange{7}{8} & \myapmb{933}{77}   & \myapmb{658}{86}   & 0.77 
                     & \myapmb{1342}{59}  & \myapmb{747}{60}   & 1.15 \\
      \myrange{8}{9} & \myapmb{660}{47}   & \myapmb{302}{51}   & 0.90 
                     & \myapmb{817}{43}   & \myapmb{395}{42}   & 0.78 \\
      \myrange{9}{10} & \myapmb{451}{34}  & \myapmb{142}{35}   & 0.82 
                      & \myapmb{501}{32}  & \myapmb{256}{31}   & 1.09 \\
      \myrange{10}{11} & \myapmb{255}{25} & \myapmb{86}{26}    & 1.13 
                       & \myapmb{317}{26} & \myapmb{188}{25}   & 0.85 \\
      \myrange{11}{12} & \myapmb{129}{28} & \myapmb{99}{30}    & 0.87 
                       & \myapmb{222}{19} & \myapmb{103}{18}   & 0.93 \\
      \myrange{12}{13} & \myapmb{129}{16} & \myapmb{46}{15}    & 1.09 
                       & \myapmb{154}{15} & \myapmb{50}{13}    & 0.98 \\
      \myrange{13}{15} & \myapmb{127}{18} & \myapmb{42}{20}    & 0.91 
                       & \myapmb{158}{18} & \myapmb{63}{17}    & 1.05\\ \hline
    \end{tabular}
  \end{center}
\end{table*}


\subsection{Efficiencies}
\label{sec:Efficiencies}

\begin{figure}
  \begin{center}
    \ifthenelse{\boolean{pdflatex}}{
      \includegraphics*[width=0.6\textwidth]{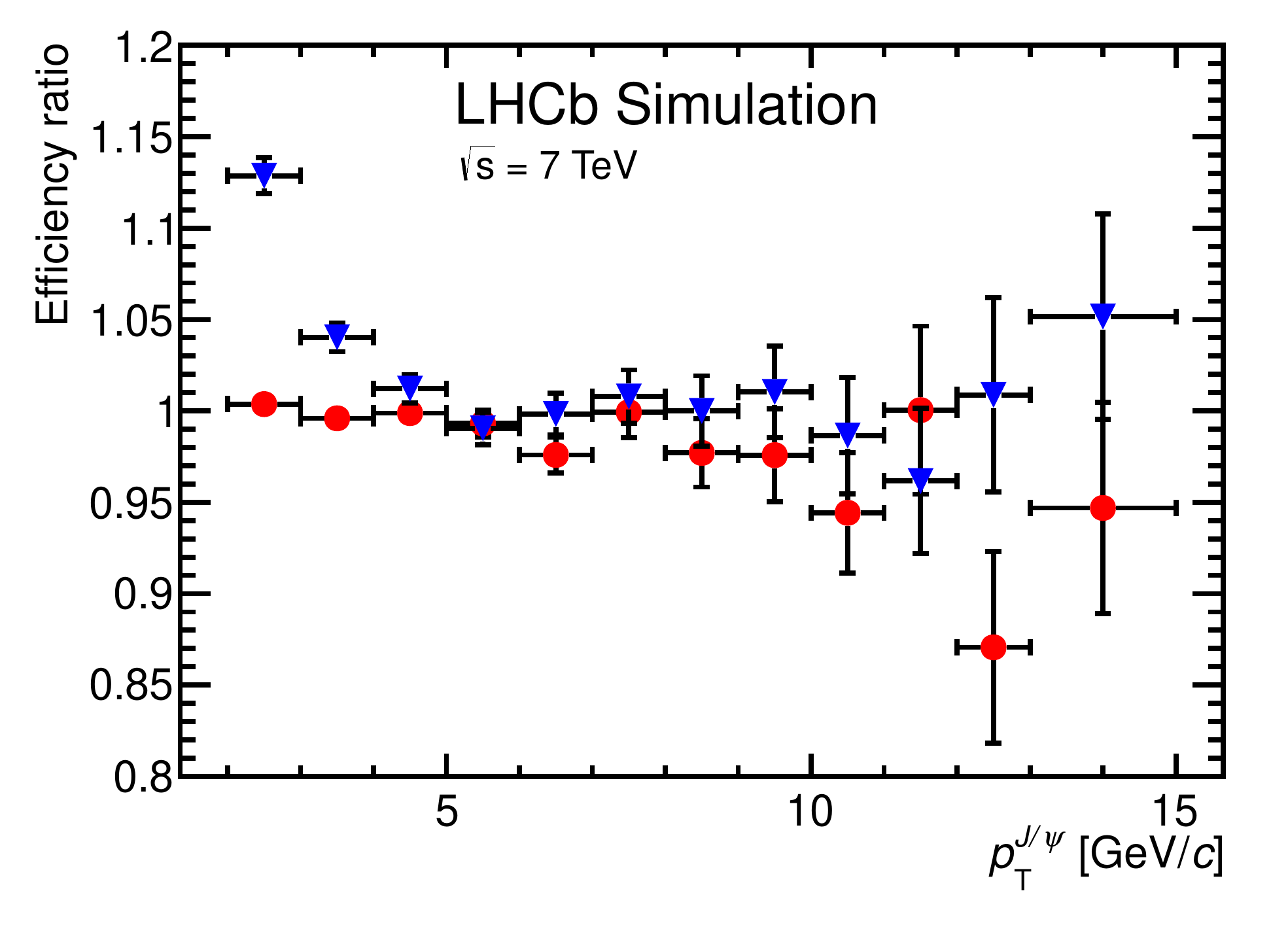}
    }{
      \includegraphics*[width=0.6\textwidth]{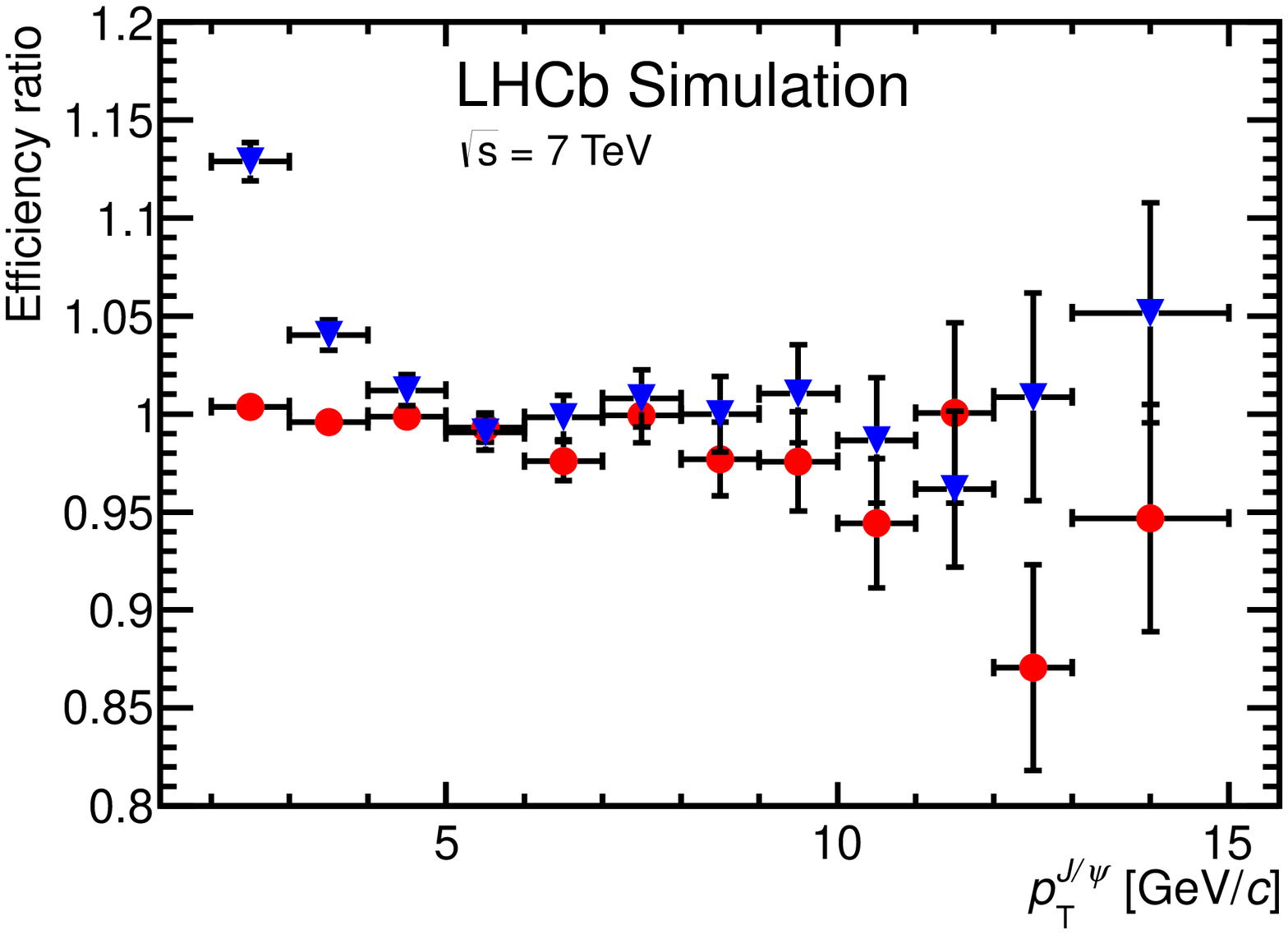}}
  \end{center}
  \caption{
    \label{fig:EffRatios}
    \small{Reconstruction and selection efficiency ratios in bins of
	\pTJpsi. The ratio of the \Jpsi\ efficiency
	(\ensuremath{\epsilon^{\ChicTwo}_{\Jpsi}\myop{/}\epsilon^{\ChicOne}_{\Jpsi}})
	is shown with red circles. The ratio of the photon
	reconstruction and selection efficiency times the \Chic\ selection
	efficiency
	(\ensuremath{\epsilon^{\ChicTwo}_{\gamma}\epsilon^{\ChicTwo}_{\mathrm{sel}}\myop{/}\epsilon^{\ChicOne}_{\gamma}\epsilon^{\ChicOne}_{\mathrm{sel}}})
	is shown with blue triangles.}}
\end{figure}

The efficiency ratios to reconstruct and select \Chic\ candidates are obtained
from simulation.  Since the photon interaction with material is not part of the
event generation procedure, the individual efficiencies for converted and
non-converted candidates are not separated. Therefore, the combined
efficiencies are calculated. The ratios of the overall efficiency for the
detection of \Jpsi\ mesons originating from the decay of a \ChicOne\ compared to
a \ChicTwo,
\ensuremath{\epsilon^{\ChicTwo}_{\Jpsi}\myop{/}\epsilon^{\ChicOne}_{\Jpsi}}, are
consistent with unity for all \pTJpsi\ bins, as shown in
\aFigure~\ref{fig:EffRatios}. The ratios of the efficiencies for reconstructing
and selecting photons from \Chic\ decays and then selecting the \Chic,
\ensuremath{\epsilon^{\ChicTwo}_{\gamma}\epsilon^{\ChicTwo}_{\mathrm{sel}}\myop{/}\epsilon^{\ChicOne}_{\gamma}\epsilon^{\ChicOne}_{\mathrm{sel}}},
are also shown in \aFigure~\ref{fig:EffRatios}. In general these efficiency
ratios are consistent with unity, except in the \pTJpsi\ bins
\myvalue{\myrange{2}{3}}{\GeVc} and \myvalue{\myrange{3}{4}}{\GeVc} where the
reconstruction and detection efficiencies for \ChicOne\ are smaller than for
\ChicTwo. The increase in the efficiency ratio in these bins arises because the
photon \pT\ spectra are different for \ChicOne\ and \ChicTwo. The photon
\ensuremath{\pTGamma\myop{>}\myvalue{650}{\MeVc}} requirement cuts harder in the
case of the \ChicOne\ and therefore lowers this efficiency. The increase in the
efficiency ratio is a kinematic effect, rather than a reconstruction effect, and
is well modelled by the simulation.


\subsection{Polarisation}
\label{sec:Polarisation}

The production of polarised \Chic\ states would modify the efficiencies
calculated from the simulation, which assumes unpolarised \Chic. 
A measurement of the \Chic\ polarisation would require an angular analysis, which is not
feasible with the present amount of data. 
Various polarisation scenarios are considered in
\aTable~\ref{tab:PolFactors}. 
Assuming no azimuthal dependence in the production process, 
the \ChicToJpsiGamma\ system is described by three angles:
\ensuremath{\theta_{\Jpsi}}, \ensuremath{\theta_{\Chic}} and
\ensuremath{\phi},
where \ensuremath{\theta_{\Jpsi}} is the angle between the directions of the  
positive muon in the \Jpsi\ rest frame 
and the \Jpsi\ in the \Chic\ rest frame,
\ensuremath{\theta_{\Chic}} is the angle between the directions of the \Jpsi\ in the
\Chic\ rest frame and the \Chic\ in the laboratory frame,
and 
\ensuremath{\phi} is the angle between the plane formed from the 
\Chic\ and \Jpsi\ momentum vectors 
in the laboratory frame and the \Jpsi\ decay plane in the \Jpsi\ rest frame.
The angular distributions are independent of the choice of 
polarisation axis (the direction of the \Chic\ in the laboratory frame)
and are detailed in \aReference~\cite{Abt:2008ed}. 
For each simulated event in the unpolarised sample, a
weight is calculated from the distribution of these angles in the various
polarisation hypotheses compared to the unpolarised distribution. 
The weights in
\aTable~\ref{tab:PolFactors} are then the average of these per-event weights in
the simulated sample. For a given (\ensuremath{|m_{\ChicOne}|},
\ensuremath{|m_{\ChicTwo}|}) polarisation combination, the central value of the
determined cross-section ratio in each \pTJpsi\ bin should be multiplied by the
number in the table. The maximum effect from the possible polarisation of the
\ChicOne\ and \ChicTwo\ mesons is given separately from the systematic
uncertainties in \aTable~\ref{tab:Results} and
\aFigure~\ref{fig:Results}.
\begin{table*}[t] 
  \small
  \renewcommand{\arraystretch}{1.8}
  \setlength{\tabcolsep}{1mm}
  \caption{
    \label{tab:PolFactors}
    \small{Polarisation weights in \pTJpsi\ bins for different combinations
      of \ChicOne\ and \ChicTwo\ polarisation states \ensuremath{|J,m_{\ChicJ}\rangle}
      with \ensuremath{|m_{\ChicJ}|=0,\cdots J}.
      The polarisation axis is defined as the direction of 
      the \Chic\ in the laboratory frame.
      Unpol. means the \Chic\ is unpolarised.}}
\medskip
  \begin{center}
    \begin{tabular}{|c|c|c|c|c|c|c|c|c|c|c|c|c|c|} \hline
      \multirow{2}{*}{(\ensuremath{|m_{\ChicOne}|,|m_{\ChicTwo}|})} & \multicolumn{12}{|c|}{\pTJpsi\ (\GeVc)} \\ \cline{2-13} 
      & \myTabRange{2}{3} & \myTabRange{3}{4} & \myTabRange{4}{5} & \myTabRange{5}{6} & \myTabRange{6}{7} & \myTabRange{7}{8} & \myTabRange{8}{9} & \myTabRange{9}{10} & \myTabRange{10}{11} & \myTabRange{11}{12} & \myTabRange{12}{13} & \myTabRange{13}{15}\\ \hline
      (Unpol,0)  & 0.99 & 0.97 & 0.94 & 0.91 & 0.88 & 0.87 & 0.86 & 0.86 & 0.86 & 0.85 & 0.85 & 0.88 \\ 
      (Unpol,1)  & 0.97 & 0.98 & 0.97 & 0.95 & 0.94 & 0.94 & 0.93 & 0.93 & 0.93 & 0.93 & 0.93 & 0.93 \\ 
      (Unpol,2)  & 1.03 & 1.04 & 1.07 & 1.11 & 1.14 & 1.17 & 1.18 & 1.18 & 1.19 & 1.18 & 1.19 & 1.16 \\ 
      (0,Unpol)  & 1.01 & 0.99 & 0.97 & 0.93 & 0.90 & 0.89 & 0.87 & 0.86 & 0.85 & 0.87 & 0.86 & 0.84 \\ 
      (1,Unpol)  & 0.99 & 1.00 & 1.02 & 1.04 & 1.05 & 1.06 & 1.06 & 1.07 & 1.08 & 1.07 & 1.07 & 1.08 \\ 
      (0,0)  & 1.00 & 0.97 & 0.91 & 0.84 & 0.80 & 0.77 & 0.75 & 0.74 & 0.72 & 0.74 & 0.74 & 0.74 \\ 
      (0,1)  & 0.98 & 0.97 & 0.93 & 0.88 & 0.85 & 0.83 & 0.81 & 0.81 & 0.79 & 0.81 & 0.81 & 0.78 \\ 
      (0,2)  & 1.04 & 1.03 & 1.03 & 1.03 & 1.03 & 1.03 & 1.03 & 1.02 & 1.00 & 1.03 & 1.03 & 0.98 \\ 
      (1,0)  & 0.99 & 0.97 & 0.96 & 0.94 & 0.93 & 0.92 & 0.92 & 0.92 & 0.92 & 0.91 & 0.91 & 0.95 \\ 
      (1,1)  & 0.97 & 0.98 & 0.98 & 0.99 & 0.99 & 0.99 & 0.99 & 1.00 & 1.00 & 1.00 & 1.00 & 1.01 \\ 
      (1,2)  & 1.03 & 1.04 & 1.09 & 1.15 & 1.20 & 1.23 & 1.26 & 1.26 & 1.28 & 1.26 & 1.27 & 1.25\\ \hline
     \end{tabular}
  \end{center}
\end{table*}


\section{Systematic uncertainties}
\label{sec:Systematics}


The branching fractions used in the analysis are
\ensuremath{\BR(\ChicOneToJpsiGamma)\myop{=}\myapmb{0.344}{0.015}} and
\ensuremath{\BR(\ChicTwoToJpsiGamma)\myop{=}\myapmb{0.195}{0.008}}, taken from
\aReference~\cite{Nakamura:2010zzi}. 
The relative systematic uncertainty on the
cross-section ratio resulting from the \ChicToJpsiGamma\ branching fractions is
6\%;
the absolute uncertainty is given for each bin of \pTJpsi\ in
\aTable~\ref{tab:Systematics}.

The simulation sample used to calculate the efficiencies has approximately the
same number of \Chic\ candidates as are observed in the data. The statistical
errors from the finite number of simulated events are included as a systematic
uncertainty in the final results. The uncertainty associated to this is
determined by sampling the efficiencies used in
\aEquation~\ref{e:ChiC1_ChiC2_ratio} according to their errors. The relative
systematic uncertainty due to the limited size of the simulation sample is found
to be in the range (\myrange{0.6}{7.2})\% and is given for each \pTJpsi\ bin in
\aTable~\ref{tab:Systematics}.

The measured \Chic\ yields depend on the values of the fixed parameters and the
fit range used. The associated systematic uncertainty has been evaluated by
repeating the fit many times, changing the values of the fixed parameters and
the fit range. Since the uncertainties arising from the fixed parameters are
expected to be correlated, a single procedure is used simultaneously varying all
these parameters. 
The \Chic\ mass
difference parameters are sampled from two Gaussian distributions with widths
taken from the errors on the masses given in
\aReference~\cite{Nakamura:2010zzi}. The mass resolution ratios,
\SigmaResRatio{\ChicTwo}{\ChicOne} and \SigmaResRatio{\ChicZero}{\ChicOne}, are
varied according to the error matrix of the fit to the simulated sample in the
range \pTJpsiRangeIntSample.

The mass resolution \SigmaRes{\ChicOne} is also determined using
a simplified background model and fitting in a reduced range. Simulation studies
show that the value of \SigmaRes{\ChicOne} also has a weak
dependence on \pTJpsi. 
The mass resolution \SigmaRes{\ChicOne} is randomly sampled from the values 
obtained 
from the default fit (described in
\aSection~\ref{sec:ExpMethod}) according to its error,
the simplified fit, again according to its error, 
and by modifying it in each
\pTJpsi\ bin according to the variation observed in the simulation.

The systematic uncertainty associated with the shape of the fitted background
function is incorporated by including or excluding the \ChicZero\ signal shape,
which peaks in the region where the background shape is most sensitive.

The background shape is also sensitive to the rise in the \DeltaM\
distribution. The systematic uncertainty from this is included by varying the
lower edge of the fit range in the interval \mypma{\myvalue{10}{\MeVcc}} around
its nominal value for each bin in \pTJpsi.

The overall systematic uncertainty from the fit is then determined from the
distribution of the \ensuremath{\ChicTwo\,/\,\ChicOne}
cross-section ratios by repeating the sampling procedure
described above many times. The relative uncertainty is found to be in the range
(\myrange{2.2}{14.6})\% and is given for each bin of \pTJpsi\ in
\aTable~\ref{tab:Systematics}.

A systematic uncertainty related to the calibration of the simulation
is evaluated by
performing the analysis on simulated events and comparing the
efficiency-corrected ratio of yields,
\ensuremath{(N_{\ChicTwo}\myop{/}N_{\ChicOne})\myop{\cdot}(\epsilon^{\ChicOne}\myop{/}{\epsilon^{\ChicTwo}})},
to the true ratio generated in the sample. A deviation of \ensuremath{-9.6\%} is
observed, caused by non-Gaussian signal shapes in the simulation from the
calorimeter calibration. These are not seen in the data, which is well described
by Gaussian signal shapes. 
The deviation is included as a systematic error,
by sampling from the negative half of a Gaussian with zero mean and a width of
\ensuremath{9.6\%}. The relative uncertainty on the cross-section ratio is
found to be less than \ensuremath{6.0\%} and is given for each bin of \pTJpsi\
in \aTable~\ref{tab:Systematics}.
A second check of the procedure was performed using simulated events generated according to 
the distributions observed in the data, 
\ie\ three overlapping Gaussians and a background shape similar to that in
\aFigure~\ref{fig:MassPlots}. 
In this case no evidence for a deviation was observed.
Other systematic uncertainties due to the modelling of the 
detector in the simulation are negligible.

In summary, the overall systematic uncertainty,
excluding that due to the branching fractions, 
is evaluated by simultaneously
sampling the deviation of the cross-section ratio from the central value, using
the distributions of the cross-section ratios described above. 
The separate systematic
uncertainties are shown in bins of \pTJpsi\ in \aTable~\ref{tab:Systematics} and
the combined uncertainties are shown in \aTable~\ref{tab:Results}.
\begin{table*}[t] \small
  \renewcommand{\arraystretch}{2}
  \caption{
    \label{tab:Systematics}
    \small{Summary of the systematic uncertainties (absolute values) on \srChicTwoToChicOne\
      in each \pTJpsi\ bin.}}
  \begin{center}
    \begin{tabular}{|c|c|c|c|c|c|c|} \hline
      \pTJpsi (\GeVc) & \myrange{2}{3} & \myrange{3}{4} & \myrange{4}{5} & \myrange{5}{6} & \myrange{6}{7} & \myrange{7}{8} \\ \hline
      Branching fractions & 
\ensuremath{{}^{+0.08}_{-0.08}} & 
\ensuremath{{}^{+0.08}_{-0.08}} & 
\ensuremath{{}^{+0.06}_{-0.06}} & 
\ensuremath{{}^{+0.07}_{-0.07}} & 
\ensuremath{{}^{+0.07}_{-0.07}} & 
\ensuremath{{}^{+0.06}_{-0.06}} \\ 
Size of simulation sample & 
\ensuremath{{}^{+0.01}_{-0.01}} & 
\ensuremath{{}^{+0.01}_{-0.01}} & 
\ensuremath{{}^{+0.01}_{-0.01}} & 
\ensuremath{{}^{+0.01}_{-0.01}} & 
\ensuremath{{}^{+0.02}_{-0.01}} & 
\ensuremath{{}^{+0.02}_{-0.02}} \\
      Fit model  & 
\ensuremath{{}^{+0.04}_{-0.05}} & 
\ensuremath{{}^{+0.05}_{-0.04}} & 
\ensuremath{{}^{+0.03}_{-0.03}} & 
\ensuremath{{}^{+0.03}_{-0.03}} & 
\ensuremath{{}^{+0.03}_{-0.04}} & 
\ensuremath{{}^{+0.05}_{-0.04}} \\
      Simulation calibration & 
\ensuremath{{}^{+0.00}_{-0.08}} & 
\ensuremath{{}^{+0.00}_{-0.07}} & 
\ensuremath{{}^{+0.00}_{-0.05}} & 
\ensuremath{{}^{+0.00}_{-0.05}} & 
\ensuremath{{}^{+0.00}_{-0.06}} & 
\ensuremath{{}^{+0.00}_{-0.06}} \\
      \hline
      \pTJpsi (\GeVc) & \myrange{8}{9} & \myrange{9}{10} & \myrange{10}{11} & \myrange{11}{12} & \myrange{12}{13} & \myrange{13}{15} \\ \hline
      Branching fractions & 
\ensuremath{{}^{+0.05}_{-0.05}} & 
\ensuremath{{}^{+0.05}_{-0.05}} & 
\ensuremath{{}^{+0.05}_{-0.05}} & 
\ensuremath{{}^{+0.06}_{-0.06}} & 
\ensuremath{{}^{+0.04}_{-0.04}} & 
\ensuremath{{}^{+0.04}_{-0.04}} \\
      Size of simulation sample & 
\ensuremath{{}^{+0.02}_{-0.02}} & 
\ensuremath{{}^{+0.02}_{-0.02}} & 
\ensuremath{{}^{+0.04}_{-0.04}} & 
\ensuremath{{}^{+0.06}_{-0.06}} & 
\ensuremath{{}^{+0.05}_{-0.05}} & 
\ensuremath{{}^{+0.05}_{-0.05}} \\
      Fit model & 
\ensuremath{{}^{+0.03}_{-0.04}} & 
\ensuremath{{}^{+0.03}_{-0.03}} & 
\ensuremath{{}^{+0.03}_{-0.03}} & 
\ensuremath{{}^{+0.02}_{-0.13}} & 
\ensuremath{{}^{+0.02}_{-0.02}} & 
\ensuremath{{}^{+0.08}_{-0.03}} \\
      Simulation calibration & 
\ensuremath{{}^{+0.00}_{-0.04}} & 
\ensuremath{{}^{+0.00}_{-0.04}} & 
\ensuremath{{}^{+0.00}_{-0.05}} & 
\ensuremath{{}^{+0.00}_{-0.06}} & 
\ensuremath{{}^{+0.00}_{-0.04}} & 
\ensuremath{{}^{+0.00}_{-0.03}} \\ \hline
    \end{tabular}
  \end{center}
\end{table*}


\section{Results and conclusions}
\label{sec:Results}

\begin{figure}[htbp]
  \begin{center}
    \subfigure{
    \ifthenelse{\boolean{pdflatex}}{
      \includegraphics*[width=0.55\textwidth]{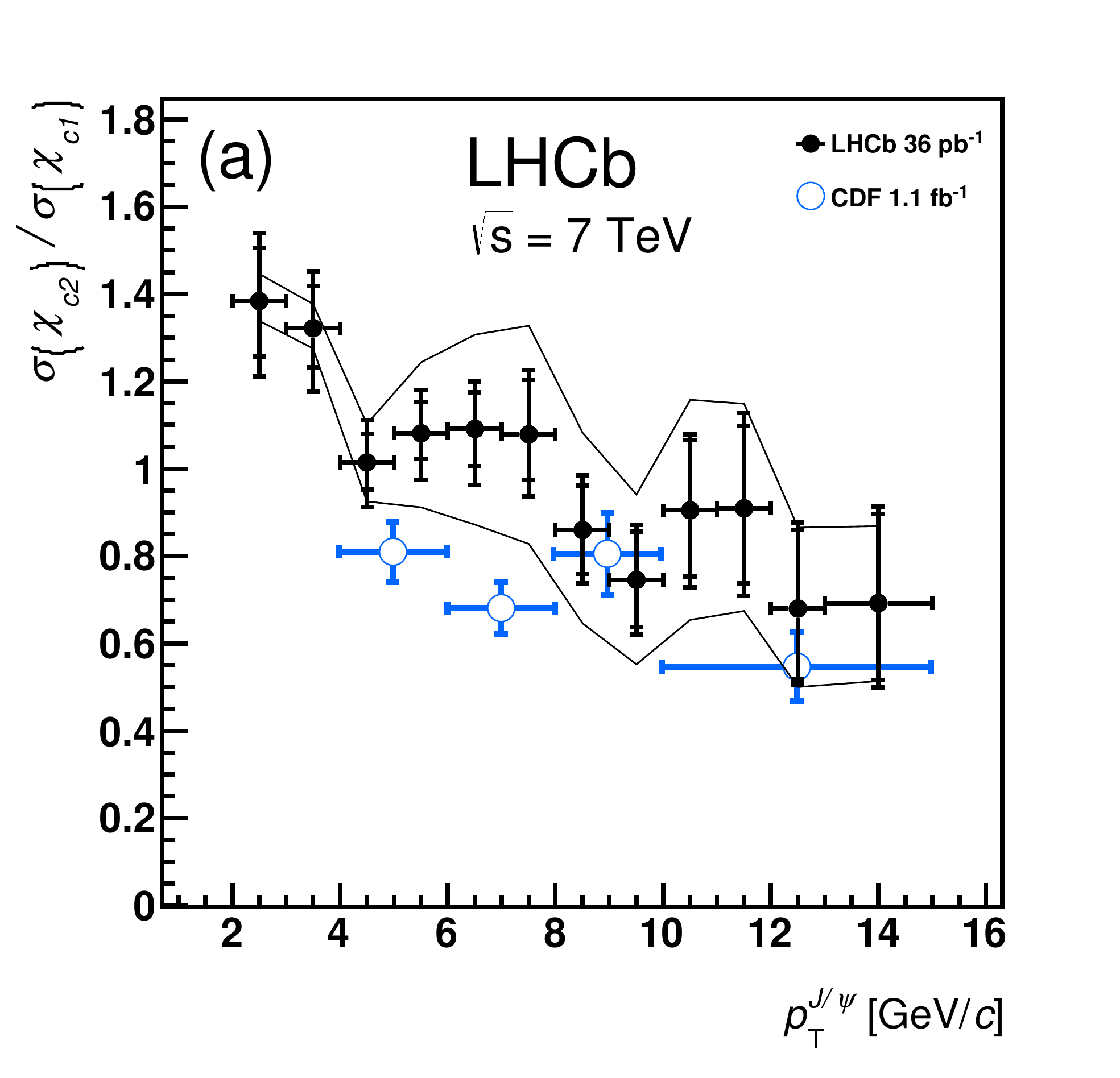}
    }{
      \includegraphics*[width=0.55\textwidth]{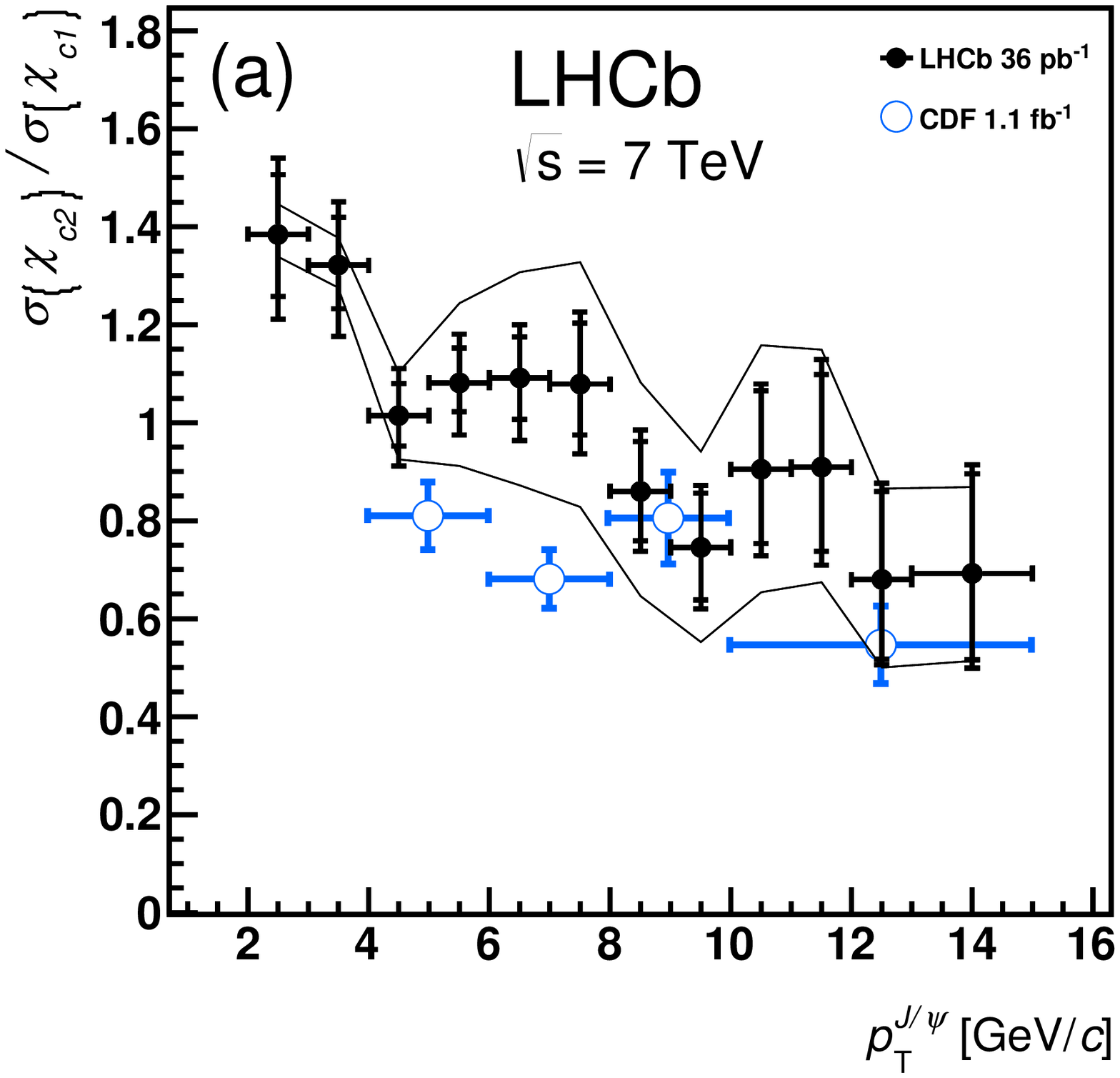}}}
    \subfigure{
    \ifthenelse{\boolean{pdflatex}}{
      \includegraphics*[width=0.55\textwidth]{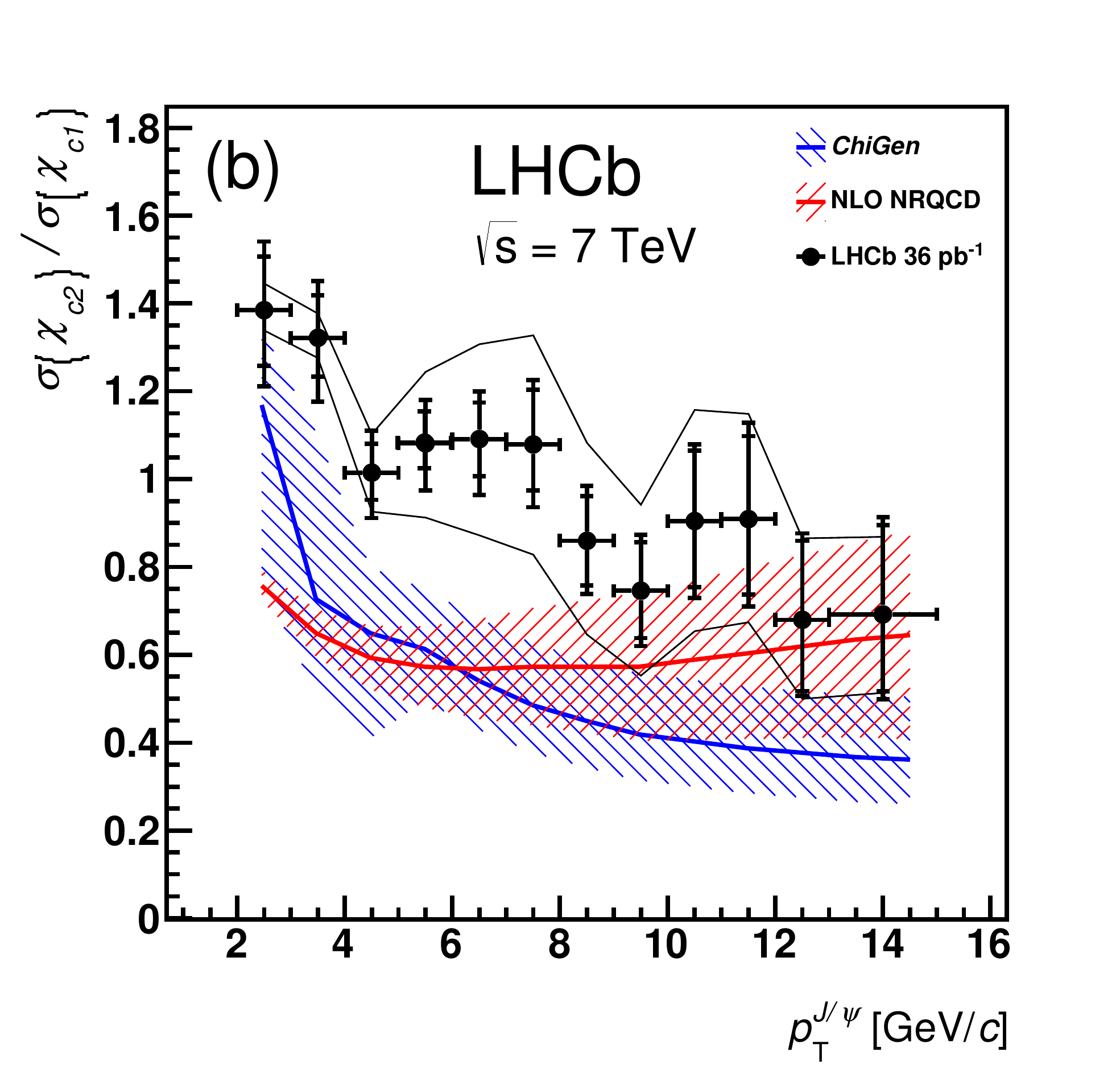}
    }{
      \includegraphics*[width=0.55\textwidth]{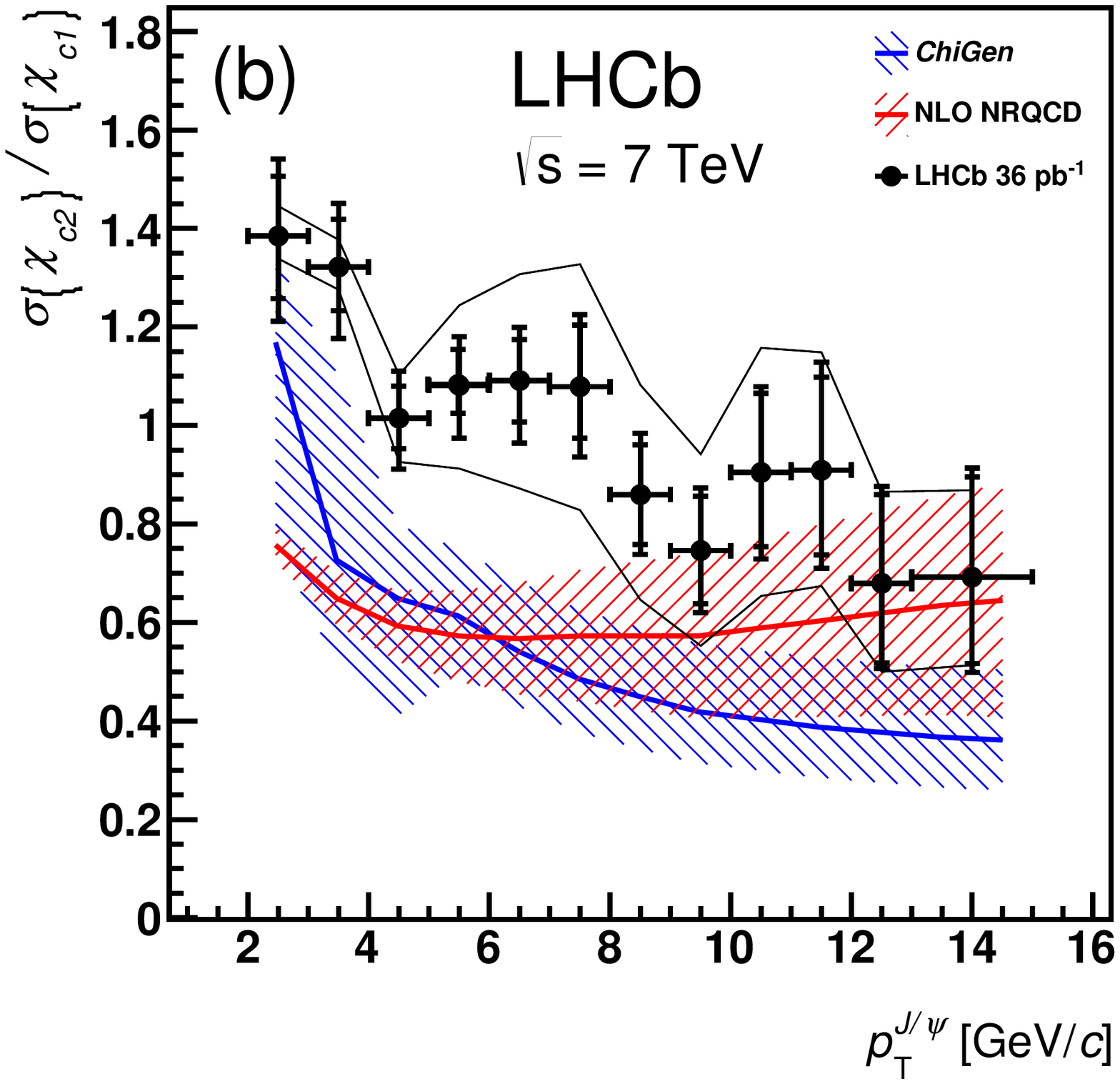}}}
  \end{center}
  \caption{\small{Ratio \srChicTwoToChicOne\ in bins of \pTJpsiRange. 
      The \LHCb\ results, in the rapidity range \yRange\ and 
      assuming the production of unpolarised \Chic\ mesons, are shown
      with solid black circles and the internal error bars correspond to the
      statistical error; the external error bars include the contribution from
      the systematic uncertainties (apart from the polarisation). The lines
      surrounding the data points show the maximum effect of the unknown \Chic\
      polarisations on the result. The upper and lower limits correspond to the
      spin states as described in the text. The CDF data points, at
      \ensuremath{\sqrt{s}\myop{=}\myvalue{1.96}{\TeV}} 
      in \ppbar\ collisions and 
      in the \Jpsi\ pseudo-rapidity range \ensuremath{|\eta^{\Jpsi}|<1.0},
      are shown in (a) with open blue circles~\cite{Abulencia:2007bra}. 
      The two hatched bands in (b) correspond
      to the \chigen\ Monte Carlo generator~\cite{Harland:ChiGen}
      and NLO NRQCD~\cite{Ma:2010vd} predictions.}}
  \label{fig:Results}
\end{figure}

The cross-section ratio, \srChicTwoToChicOne, measured in bins of \pTJpsi\ is
given in \aTable~\ref{tab:Results} and shown in \aFigure~\ref{fig:Results}. 
Previous measurements from WA11 in \ensuremath{\pim}Be collisions at \myvalue{185}{\GeVc} gave
\ensuremath{\srChicTwoToChicOne\myop{=}\myapmb{1.4}{0.6}}~\cite{Lemoigne:1982jc},
and from HERA-B in \pA\ collisions at
\ensuremath{\sqrt{s}\myop{=}\myvalue{41.6}{\GeV}} with
\pTJpsi\ below roughly \myvalue{5}{\GeVc}
gave
\ensuremath{\srChicTwoToChicOne\myop{=}\myapmb{1.75}{0.7}}~\cite{Abt:2008ed}.  
The data points from CDF~\cite{Abulencia:2007bra} at
\ensuremath{\sqrt{s}\myop{=}\myvalue{1.96}{\TeV}} in \ppbar\ collisions are also
shown in \aFigure~\ref{fig:Results}a).
\begin{table}[t]
  \renewcommand{\arraystretch}{2}
  \caption{
    \label{tab:Results}
    \small{Ratio \srChicTwoToChicOne\ in bins of \pTJpsi\ in the range
      \pTJpsiRange\ and in the
      rapidity range \yRange. The first error is the statistical error, the
      second is the systematic uncertainty (apart from the branching fraction and
      polarisation) and the third is due to the \ChicToJpsiGamma\ branching
      fractions. Also given is the maximum effect of the unknown \Chic\
      polarisations on the result as described in \aSection~\ref{sec:Polarisation}.}}
\medskip
  \begin{center}
    \begin{tabular}{|c|c|c|} \hline
      \pTJpsi\ (\GeVc) & \srChicTwoToChicOne & Polarisation effects \\ \hline
      \myrange{2}{3} & \ensuremath{{1.39}^{+0.12\;+0.06\;+0.08}_{-0.13\;-0.09\;-0.08}} & \ensuremath{{}^{+0.06}_{-0.05}}\\ \hline
      \myrange{3}{4} & \ensuremath{{1.32}^{+0.10\;+0.03\;+0.08}_{-0.09\;-0.09\;-0.08}} & \ensuremath{{}^{+0.06}_{-0.05}}\\ \hline
      \myrange{4}{5} & \ensuremath{{1.02}^{+0.07\;+0.04\;+0.06}_{-0.06\;-0.06\;-0.06}} & \ensuremath{{}^{+0.09}_{-0.09}}\\ \hline
      \myrange{5}{6} & \ensuremath{{1.08}^{+0.07\;+0.04\;+0.07}_{-0.06\;-0.06\;-0.07}} & \ensuremath{{}^{+0.16}_{-0.17}}\\ \hline
      \myrange{6}{7} & \ensuremath{{1.09}^{+0.08\;+0.03\;+0.07}_{-0.09\;-0.07\;-0.07}} & \ensuremath{{}^{+0.22}_{-0.22}}\\ \hline
      \myrange{7}{8} & \ensuremath{{1.08}^{+0.13\;+0.05\;+0.06}_{-0.10\;-0.07\;-0.06}} & \ensuremath{{}^{+0.25}_{-0.25}}\\ \hline
      \myrange{8}{9} & \ensuremath{{0.86}^{+0.10\;+0.04\;+0.05}_{-0.10\;-0.06\;-0.05}} & \ensuremath{{}^{+0.22}_{-0.21}}\\ \hline
      \myrange{9}{10} & \ensuremath{{0.75}^{+0.11\;+0.04\;+0.05}_{-0.11\;-0.06\;-0.05}} & \ensuremath{{}^{+0.20}_{-0.19}}\\ \hline
      \myrange{10}{11} & \ensuremath{{0.91}^{+0.16\;+0.05\;+0.05}_{-0.15\;-0.07\;-0.05}} & \ensuremath{{}^{+0.25}_{-0.25}}\\ \hline
      \myrange{11}{12} & \ensuremath{{0.91}^{+0.19\;+0.09\;+0.06}_{-0.17\;-0.10\;-0.06}} & \ensuremath{{}^{+0.24}_{-0.24}}\\ \hline
      \myrange{12}{13} & \ensuremath{{0.68}^{+0.18\;+0.05\;+0.04}_{-0.16\;-0.07\;-0.04}} & \ensuremath{{}^{+0.19}_{-0.18}}\\ \hline
      \myrange{13}{15} & \ensuremath{{0.69}^{+0.20\;+0.07\;+0.04}_{-0.18\;-0.07\;-0.04}} & \ensuremath{{}^{+0.18}_{-0.18}}\\ \hline
    \end{tabular}
  \end{center}
\end{table}

Theoretical predictions,
calculated in the LHCb rapidity range \yRange,
from the \chigen\ Monte Carlo
generator~\cite{Harland:ChiGen},
which is an implementation of the
leading-order colour-singlet model described in \aReference~\cite{Glover:1987az},
and from the NLO NRQCD calculations~\cite{Ma:2010vd}
are shown in \aFigure~\ref{fig:Results}b). 
The hatched bands represent the uncertainties in the theoretical
predictions.

\Figure~\ref{fig:Results} also shows the maximum effect of the unknown \Chic\
polarisations on the result, shown as the lines surrounding the data
points. In the first \pTJpsi\ bin, the upper limit 
corresponds to the spin state combination
\ensuremath{(|m_{\ChicOne}|, |m_{\ChicTwo}|)\myop{=}(0,2)} and the
lower limit corresponds to the spin state combination \ensuremath{(1,1)}. In all
subsequent \pTJpsi\ bins, the upper limit corresponds to spin state combination
\ensuremath{(1,2)} and the lower limit corresponds to \ensuremath{(0,0)}.

In summary, the ratio of the \srChicTwoToChicOne\ prompt production
cross-sections has been measured as a function of \pTJpsi\
using \myvalue{36}{\invpb} of data
collected by \LHCb\ during 2010 at a centre-of-mass energy \SqrtS.
The \chigen\ generator describes the shape of the distribution reasonably
well, although the data lie consistently above the model prediction. 
This could be explained by important higher order perturbative corrections and/or sizeable
colour octet terms not included in the calculation. 
The results are in agreement with the NLO NRQCD model for
\ensuremath{\pTJpsi\myop{>}\myvalue{8}{\GeVc}}.


\section*{Acknowledgments}
We would like to thank L.~A.~Harland-Lang, W.~J.~Stirling and K.~Chao for
supplying the theory predictions for comparison 
to our data and for many helpful discussions.

We express our gratitude to our colleagues in the CERN accelerator
departments for the excellent performance of the LHC. We thank the
technical and administrative staff at CERN and at the LHCb institutes,
and acknowledge support from the National Agencies: CAPES, CNPq,
FAPERJ and FINEP (Brazil); CERN; NSFC (China); CNRS/IN2P3 (France);
BMBF, DFG, HGF and MPG (Germany); SFI (Ireland); INFN (Italy); FOM and
NWO (The Netherlands); SCSR (Poland); ANCS (Romania); MinES of Russia and
Rosatom (Russia); MICINN, XuntaGal and GENCAT (Spain); SNSF and SER
(Switzerland); NAS Ukraine (Ukraine); STFC (United Kingdom); NSF
(USA). We also acknowledge the support received from the ERC under FP7
and the Region Auvergne.


\bibliographystyle{LHCb}
\bibliography{main}

\ifx\mcitethebibliography\mciteundefinedmacro
\PackageError{LHCb.bst}{mciteplus.sty has not been loaded}
{This bibstyle requires the use of the mciteplus package.}\fi
\providecommand{\href}[2]{#2}
\begin{mcitethebibliography}{10}
\mciteSetBstSublistMode{n}
\mciteSetBstMaxWidthForm{subitem}{\alph{mcitesubitemcount})}
\mciteSetBstSublistLabelBeginEnd{\mcitemaxwidthsubitemform\space}
{\relax}{\relax}

\bibitem{Bodwin:1994jh}
G.~T. Bodwin, E.~Braaten, and G.~Lepage,
  \ifthenelse{\boolean{articletitles}}{{\it {Rigorous QCD analysis of inclusive
  annihilation and production of heavy quarkonium}}, }{}
  \href{http://dx.doi.org/10.1103/PhysRevD.51.1125, 10.1103/PhysRevD.55.5853}{
  {\em Phys. Rev.} {\bf D51} (1995) 1125},
  [\href{http://xxx.lanl.gov/abs/hep-ph/9407339}{{\tt arXiv:hep-ph/9407339}}].
  Erratum-ibid. D55 (1997) 5853\relax
\mciteBstWouldAddEndPuncttrue
\mciteSetBstMidEndSepPunct{\mcitedefaultmidpunct}
{\mcitedefaultendpunct}{\mcitedefaultseppunct}\relax
\EndOfBibitem
\bibitem{Campbell:2007ws}
J.~M. Campbell, F.~Maltoni, and F.~Tramontano,
  \ifthenelse{\boolean{articletitles}}{{\it {QCD corrections to $J/\psi$ and
  Upsilon production at hadron colliders}}, }{}
  \href{http://dx.doi.org/10.1103/PhysRevLett.98.252002}{ {\em Phys. Rev.
  Lett.} {\bf 98} (2007) 252002},
  [\href{http://xxx.lanl.gov/abs/hep-ph/0703113}{{\tt
  arXiv:hep-ph/0703113}}]\relax
\mciteBstWouldAddEndPuncttrue
\mciteSetBstMidEndSepPunct{\mcitedefaultmidpunct}
{\mcitedefaultendpunct}{\mcitedefaultseppunct}\relax
\EndOfBibitem
\bibitem{Ma:2010vd}
Y.-Q. Ma, K.~Wang, and K.-T. Chao, \ifthenelse{\boolean{articletitles}}{{\it
  {QCD radiative corrections to $\chi_{cJ}$ production at hadron colliders}},
  }{} \href{http://dx.doi.org/10.1103/PhysRevD.83.111503}{ {\em Phys. Rev.}
  {\bf D83} (2011) 111503}, [\href{http://xxx.lanl.gov/abs/1002.3987}{{\tt
  arXiv:1002.3987}}]\relax
\mciteBstWouldAddEndPuncttrue
\mciteSetBstMidEndSepPunct{\mcitedefaultmidpunct}
{\mcitedefaultendpunct}{\mcitedefaultseppunct}\relax
\EndOfBibitem
\bibitem{Lemoigne:1982jc}
WA11 collaboration, Y.~Lemoigne et~al.,
  \ifthenelse{\boolean{articletitles}}{{\it {Measurement of hadronic production
  of the $\chi_{1}^{++}(3507)$ and the $\chi_{2}^{++}(3553)$ through their
  radiative decay to $J/\psi$}}, }{}
  \href{http://dx.doi.org/10.1016/0370-2693(82)90795-X}{ {\em Phys. Lett.} {\bf
  B113} (1982) 509}. Erratum-ibid. B116 (1982) 470\relax
\mciteBstWouldAddEndPuncttrue
\mciteSetBstMidEndSepPunct{\mcitedefaultmidpunct}
{\mcitedefaultendpunct}{\mcitedefaultseppunct}\relax
\EndOfBibitem
\bibitem{Abt:2008ed}
HERA-B collaboration, I.~Abt et~al., \ifthenelse{\boolean{articletitles}}{{\it
  {Production of the charmonium states $\chi_{c1}$ and $\chi_{c2}$ in proton
  nucleus interactions at $\sqrt{s}$ = 41.6 GeV}}, }{}
  \href{http://dx.doi.org/10.1103/PhysRevD.79.012001}{ {\em Phys. Rev.} {\bf
  D79} (2009) 012001}, [\href{http://xxx.lanl.gov/abs/0807.2167}{{\tt
  arXiv:0807.2167}}]\relax
\mciteBstWouldAddEndPuncttrue
\mciteSetBstMidEndSepPunct{\mcitedefaultmidpunct}
{\mcitedefaultendpunct}{\mcitedefaultseppunct}\relax
\EndOfBibitem
\bibitem{Abulencia:2007bra}
CDF collaboration, A.~Abulencia et~al.,
  \ifthenelse{\boolean{articletitles}}{{\it {Measurement of
  $\sigma_{\chi_{c2}}{\cal B}(\chi_{c2} \to J/\psi \gamma)/\sigma_{\chi_{c1}}
  {\cal B}(\chi_{c1} \to J/\psi \gamma)$ in $p \bar{p}$ collisions at
  $\sqrt{s}$ = 1.96 TeV}}, }{}
  \href{http://dx.doi.org/10.1103/PhysRevLett.98.232001}{ {\em Phys. Rev.
  Lett.} {\bf 98} (2007) 232001},
  [\href{http://xxx.lanl.gov/abs/hep-ex/0703028}{{\tt
  arXiv:hep-ex/0703028}}]\relax
\mciteBstWouldAddEndPuncttrue
\mciteSetBstMidEndSepPunct{\mcitedefaultmidpunct}
{\mcitedefaultendpunct}{\mcitedefaultseppunct}\relax
\EndOfBibitem
\bibitem{Alves:2008zz}
LHCb collaboration, A.~A. Alves~Jr et~al.,
  \ifthenelse{\boolean{articletitles}}{{\it {The LHCb detector at the LHC}},
  }{} \href{http://dx.doi.org/10.1088/1748-0221/3/08/S08005}{ {\em JINST} {\bf
  3} (2008) S08005}\relax
\mciteBstWouldAddEndPuncttrue
\mciteSetBstMidEndSepPunct{\mcitedefaultmidpunct}
{\mcitedefaultendpunct}{\mcitedefaultseppunct}\relax
\EndOfBibitem
\bibitem{Sjostrand:2006za}
T.~{Sj\"{o}strand}, S.~Mrenna, and P.~Z. Skands,
  \ifthenelse{\boolean{articletitles}}{{\it {PYTHIA 6.4 physics and manual}},
  }{} \href{http://dx.doi.org/10.1088/1126-6708/2006/05/026}{ {\em JHEP} {\bf
  0605} (2006) 026}, [\href{http://xxx.lanl.gov/abs/hep-ph/0603175}{{\tt
  arXiv:hep-ph/0603175}}]\relax
\mciteBstWouldAddEndPuncttrue
\mciteSetBstMidEndSepPunct{\mcitedefaultmidpunct}
{\mcitedefaultendpunct}{\mcitedefaultseppunct}\relax
\EndOfBibitem
\bibitem{LHCb-PROC-2010-056}
I.~Belyaev et~al., \ifthenelse{\boolean{articletitles}}{{\it {Handling of the
  generation of primary events in \gauss, the \lhcb simulation framework}}, }{}
  \href{http://dx.doi.org/10.1109/NSSMIC.2010.5873949}{ {\em Nuclear Science
  Symposium Conference Record (NSS/MIC)} {\bf IEEE} (2010) 1155}\relax
\mciteBstWouldAddEndPuncttrue
\mciteSetBstMidEndSepPunct{\mcitedefaultmidpunct}
{\mcitedefaultendpunct}{\mcitedefaultseppunct}\relax
\EndOfBibitem
\bibitem{Lange:2001uf}
D.~J. Lange, \ifthenelse{\boolean{articletitles}}{{\it {The EvtGen particle
  decay simulation package}}, }{}
  \href{http://dx.doi.org/10.1016/S0168-9002(01)00089-4}{ {\em Nucl. Instrum.
  Meth.} {\bf A462} (2001) 152}\relax
\mciteBstWouldAddEndPuncttrue
\mciteSetBstMidEndSepPunct{\mcitedefaultmidpunct}
{\mcitedefaultendpunct}{\mcitedefaultseppunct}\relax
\EndOfBibitem
\bibitem{Barberio:1993qi}
E.~Barberio and Z.~W\c{a}s, \ifthenelse{\boolean{articletitles}}{{\it {\photos:
  a universal Monte Carlo for QED radiative corrections: version 2.0}}, }{}
  \href{http://dx.doi.org/10.1016/0010-4655(94)90074-4}{ {\em Comput. Phys.
  Commun.} {\bf 79} (1994) 291}\relax
\mciteBstWouldAddEndPuncttrue
\mciteSetBstMidEndSepPunct{\mcitedefaultmidpunct}
{\mcitedefaultendpunct}{\mcitedefaultseppunct}\relax
\EndOfBibitem
\bibitem{Agostinelli:2002hh}
S.~Agostinelli et~al., \ifthenelse{\boolean{articletitles}}{{\it {\geant: a
  simulation toolkit}}, }{}
  \href{http://dx.doi.org/10.1016/S0168-9002(03)01368-8}{ {\em Nucl. Instrum.
  Meth.} {\bf A506} (2003) 250}\relax
\mciteBstWouldAddEndPuncttrue
\mciteSetBstMidEndSepPunct{\mcitedefaultmidpunct}
{\mcitedefaultendpunct}{\mcitedefaultseppunct}\relax
\EndOfBibitem
\bibitem{Aaij:2011jh}
LHCb collaboration, R.~Aaij et~al., \ifthenelse{\boolean{articletitles}}{{\it
  {Measurement of $J/\psi$ production in pp collisions at $\sqrt{s}$=7 TeV}},
  }{} \href{http://dx.doi.org/10.1140/epjc/s10052-011-1645-y}{ {\em Eur. Phys.
  J.} {\bf C71} (2011) 1645}, [\href{http://xxx.lanl.gov/abs/1103.0423}{{\tt
  arXiv:1103.0423}}]\relax
\mciteBstWouldAddEndPuncttrue
\mciteSetBstMidEndSepPunct{\mcitedefaultmidpunct}
{\mcitedefaultendpunct}{\mcitedefaultseppunct}\relax
\EndOfBibitem
\bibitem{Nakamura:2010zzi}
Particle Data Group, K.~Nakamura et~al.,
  \ifthenelse{\boolean{articletitles}}{{\it {Review of particle physics}}, }{}
  \href{http://dx.doi.org/10.1088/0954-3899/37/7A/075021}{ {\em J. Phys.} {\bf
  G37} (2010) 075021}. {Includes 2011 partial update for the 2012
  edition}\relax
\mciteBstWouldAddEndPuncttrue
\mciteSetBstMidEndSepPunct{\mcitedefaultmidpunct}
{\mcitedefaultendpunct}{\mcitedefaultseppunct}\relax
\EndOfBibitem
\bibitem{Harland:ChiGen}
L.~A. Harland-Lang and W.~J. Stirling, {\em
  http://projects.hepforge.org/superchic/chigen.html}\relax
\mciteBstWouldAddEndPuncttrue
\mciteSetBstMidEndSepPunct{\mcitedefaultmidpunct}
{\mcitedefaultendpunct}{\mcitedefaultseppunct}\relax
\EndOfBibitem
\bibitem{Glover:1987az}
E.~W.~N. Glover, A.~D. Martin, and W.~J. Stirling,
  \ifthenelse{\boolean{articletitles}}{{\it {\Jpsi\ production at large
  transverse momentum at hadron colliders}}, }{}
  \href{http://dx.doi.org/10.1007/BF01584398}{ {\em Z. Phys.} {\bf C38} (1988)
  473}. Erratum-ibid. C49 (1991) 526\relax
\mciteBstWouldAddEndPuncttrue
\mciteSetBstMidEndSepPunct{\mcitedefaultmidpunct}
{\mcitedefaultendpunct}{\mcitedefaultseppunct}\relax
\EndOfBibitem
\end{mcitethebibliography}

\end{document}